\documentstyle[preprint,eqsecnum,aps,tighten]{revtex}

\begin{document}

\draft

\preprint{SNUTP-94/13 (Revised)}

\title{Heavy Quark Symmetry and the Skyrme Model}

\author{Yongseok Oh}
\address{Department of Physics, National Taiwan University,
     Taipei, Taiwan 10764, R.O.C.}
\author{Byung-Yoon Park}
\address{Department of Physics, Chungnam National University,
     Daejeon 305--764, Korea}
\author{Dong-Pil Min}
\address{Center for Theoretical Physics and Department of Physics,
     Seoul National University, Seoul 151--742, Korea}

\maketitle

\begin{abstract}
We present a consistent way of describing heavy baryons containing
a heavy quark as bound states of an $SU(2)$ soliton and heavy mesons.
The resulting mass formula reveals the heavy quark symmetry
explicitly. By extending the model to the orbitally excited states,
we establish the generic structure of the heavy baryon spectrum.
As anticipated from the heavy quark spin symmetry, the $c$-factor
denoting the hyperfine splitting constant {\em vanishes} and the
baryons with the same angular momentum of light degrees of freedom
form degenerate doublets. This approach is also applied to the
pentaquark exotic baryons, where the conventional $c$-factor plays
no more a role of the hyperfine constant. After diagonalizing the
Hamiltonian of order $N_c^{-1}$, we get the degenerate doublets,
which implies the vanishing of genuine hyperfine splitting.
\end{abstract}

\pacs{PACS number(s): 12.39.Dc, 12.39.Hg, 14.20.Lq, 14.20.Mr}


\section{Introduction}

Hadrons containing a single heavy quark ($Q$) with its mass ($m_Q$)
much greater than a typical scale of strong interactions
($\Lambda_{\mbox{\small QCD}}$) can be viewed as a freely propagating
point-like color source dressed by light degrees of freedom such as
light quarks and gluons. Besides the chiral symmetry for the light
quark system, such a system reveals an additional symmetry, so-called
the heavy quark spin-flavor symmetry\cite{IW89,HQS,Ge92} as the heavy
quark mass goes to infinity. In this limit, the heavy quark spin
decouples to the rest of the strongly interacting light quark system,
since their coupling is a relativistic effect of order $1/m_Q$.
Furthermore, since the heavy quark can hardly change its velocity
due to the strong interactions via soft gluons, the dynamics of the
system is independent of its mass and, therefore, its flavor.
As a consequence of the heavy quark spin symmetry, the hadrons come
in degenerate doublets\cite{IW91} with total spin
(regardless of the number of light quarks)
\begin{equation}
j_\pm = j_\ell \pm \textstyle\frac12 ,
\label{jpm}
\end{equation}
(unless $j_\ell=0$) which are formed by combining the spin of the
heavy quark with the total angular momentum of the light degrees of
freedom $j_\ell$. In other words, the total angular momentum of the
light degrees of freedom, ${\vec J}_\ell = {\vec J} - {\vec S}_Q$,
is conserved and the corresponding quantum number $j_\ell$ can
classify the hadrons with a single heavy quark.

In the Skyrme model {\it \`a la\/} Callan and Klebanov (CK)\cite{CK},
heavy baryons can be described by bound states of a soliton of the
$SU(2)$ chiral Lagrangian and the heavy meson containing the heavy
quark. This picture which was originally suggested for the strange
hyperons has been shown to work successfully in describing the
static properties of the heavy baryons with charm(c)- or
bottom(b)-quark\cite{RRS2,BSM,OMRS}. It is also extended to the
study of the exotic pentaquark-baryons ($P$-baryons, in short)
as the bound states of soliton and antiflavored heavy mesons\cite{RS93}.
(Here, by ``antiflavored" heavy mesons, we denote the $\bar{Q}q$
heavy mesons carrying opposite heavy flavor, {\it i.e.\,}, $C=-1$
or $B=+1$. They are the antiparticles of normal heavy mesons consist
of $Q\bar{q}$, which we will call as heavy mesons.)
However, one should be careful in applying the bound state approach
to the heavy baryons, especially in case of $P$-baryons, due to
following respects:

Firstly, one should be more careful on the interpretation and the
calculation of the so-called hyperfine constant:
When quantized, the bound system of the soliton and heavy mesons
with energy $\omega_B$ and grand spin quantum number $k$ (the
corresponding operator being defined as $\vec K = {\vec I}_h + {\vec
J}_h$; ${\vec I}_h$: isospin of heavy mesons, ${\vec J}_h$: spin of
heavy mesons) can be identified with the heavy baryon of isospin $i$
and spin $j$. The masses of such heavy baryons in the conventional
approach\cite{RRS2} turns out to be
\begin{equation}
m_{(i,j)}^{} = M_{sol} + \omega_B^{}
+ \frac{1}{2{\cal I}}\{cj(j+1) + (1-c)i(i+1) + c(c-1)k(k+1)\},
\label{MF}
\end{equation}
where $M_{sol}$ and ${\cal I}$ are the soliton mass and its moment
of inertia with respect to the collective isospin rotation,
respectively. The spin of the heavy baryon $j$ takes a priori one
of the values $|i-k|,\cdots, i+k$ and $c$, ``hyperfine constant",
is a constant defined through
\begin{equation}
\langle k,k'_3| {\vec\Theta} |k,k_3\rangle \equiv
- c \langle k,k'_3 | \vec{K} |k,k_3 \rangle,
\label{cfactor}
\end{equation}
on the analogy of the Lande's g-factor in atomic physics. Here,
${\vec\Theta}$ is the meson field operator induced by the collective
rotation, which forms the first rank tensor in the space of the
grand spin eigenstates $|k,k_3\rangle$. The $c$ yields the hyperfine
splittings between the heavy baryon masses and it has been known to
play the role of an {\it order parameter} for the heavy quark
symmetry; it is required to vanish in the heavy meson mass limit so
that the heavy baryon masses do not depend on the total {\it spin}.
{\em This requirement for the $c$, however, has been over-emphasized.}
In fact, the heavy quark symmetry does not necessarily require such an
entire independence of mass on the {\em total spin}, as will be
discussed in this work.

To heavy baryons carrying a heavy quark one may extend straightforwardly
the applicability of Eq.(\ref{MF}). The heavy quark symmetry is
manifested by vanishing $c$-factors of bound heavy meson states,
provided a subtle point is corrected\cite{OPM}. In the strangeness
sector the vector mesons $K^*$ are eliminated out in favor of the
pseudoscalar meson $K$, in analogy to the $\rho$ mesons for the
light quark system.\cite{SMNR} This approximation is valid only
when the vector meson masses are sufficiently larger than those
of the corresponding pseudoscalar mesons. As the heavy quark
mass in both mesons increases, the heavy vector meson and the
heavy pseudoscalar meson should be treated on the same footing.
Defects in taking the conventional bound state approach of CK
to heavy baryons have been pointed out and correct heavy baryon mass
spectra with the explicit heavy quark symmetry have been obtained
by Jenkins {\it et al.\/}\cite{JMW} and other
groups\cite{HB,NRZ,LNRZ,GMSS} in the infinitely heavy meson mass limit.
In Refs.\cite{JMW,HB}, a different but equivalent quantization scheme
is adopted: the soliton is first quantized to nucleons and $\Delta$'s
and then the heavy mesons are bound to them to form a heavy baryon,
while in conventional bound state approach\cite{CK} the whole
soliton-heavy meson bound system is quantized by using the collective
coordinates. The mass predictions\cite{NRZ,LNRZ} based on Eq.(\ref{MF}),
however, give slightly different results from those of Ref.\cite{JMW},
{\it i.e.\/}, constant shifts of heavy baryon masses to the amount
of $3/8{\cal I}$. This difference comes from the approximate
treatment on the last term of Eq.(\ref{MF}). We will show that the
two predictions are equivalent by treating the term in proper way.

We have met a completely different situation in the study of
penta-quark heavy baryons which carry one heavy anti-quark.\cite{OPM0}
The $c$-factors associated with the bound states of the antiflavored
heavy mesons do not vanish. In some specific cases, one gets
involved with a serious problem such as {\em negative $c$-factors},
with which the mass formula (\ref{MF}) would yield lower mass for the
baryon with higher spin. However, this flaw can be cured by the
following observation. There appear multiply degenerate heavy meson
bound states, which prevent us from using the mass formula (\ref{MF})
in the present form. The $c$-factor defined through Eq.(\ref{cfactor})
cannot play the role of the hyperfine constant. Therefore, the
hyperfine splitting being of order $1/N_c$ should be obtained, due
to the appearance of the off-diagonal terms, by diagonalizing the
Hamiltonian matrix with respect to the soliton-antiflavored heavy
meson bound states degenerate up to order $N_c^0$. We shall see
that the mixing of the states at the $N_c^{-1}$ order cannot be
neglected in anti-flavor case. The resulting $P$-baryon masses do
respect the heavy quark symmetry and {\em the resulting hyperfine
constants vanish.}

The main purpose of this paper is to clarify the things associated
with the hyperfine constant in the conventional bound state approach
and develop a consistent bound state approach to be applied not only
to the normal heavy baryons but also to negative parity heavy baryons
and exotic baryons carrying an antiflavor. It also supplies some of
the details left out in our previous paper\cite{OPM0} where we
investigated the pentaquark exotic baryons. We will work in an extreme
limit where both the soliton and heavy mesons are infinitely heavy and
sit on the same point in space. This approximation enables us to get
useful informations without getting involved with any complicated
numerical calculations.

This paper is organized as follows. In Sec.~II, we briefly introduce
our working Lagrangian density. The positive and negative parity
eigenstates of the heavy mesons under the static potentials provided
by the soliton configuration are found in Sec.~III. Section~IV is to
discuss the collective coordinate quantization procedure for the
bound system of a single heavy meson to the soliton. In Sec.~V we
describe the heavy baryons as the bound heavy mesons to soliton.
We also derive a mass formula for the heavy baryons, which is more
appropriate to appreciate the heavy quark symmetry than Eq.(\ref{MF}).
The realization of heavy quark symmetry in heavy baryon spectrum is
also discussed. We study the pentaquark exotic baryons by considering
the bound states of the ``anti-flavored" heavy mesons in Sec.~VI.
We shall show that there exist degenerate doublets as given in
Eq.(\ref{jpm}) in pentaquark states. A few concluding remarks are
given in Sec. VII and explicit formulas are provided in
Appendices~A and B.


\section{Heavy Meson Effective Lagrangian}

We start with describing briefly the effective Lagrangian for the
heavy mesons interacting with Goldstone bosons, which respects
both heavy quark symmetry and chiral
$SU(2)_L\times SU(2)_R$\footnote{We will work with two light
flavors. For the generalization to three flavors, see Ref.\cite{SU3}.}
symmetry. (See Refs.\cite{HQET,Yan,vm_HQET} for details.)

Consider heavy mesons containing a heavy quark $Q$ and a light
antiquark $\bar{q}$. Here, the light antiquark in a heavy meson is
assumed to form a point-like object with the heavy quark, endowing it
with appropriate color, flavor, angular momentum and parity.
Let $\Phi$ and $\Phi_\mu^*$ be the field operators that annihilate
$j^\pi$=$0^-$ and $1^-$ heavy mesons with $C=+1$ or $B=-1$.
These fields $\Phi$ and $\Phi^*$ form an $SU(2)$ antidoublets:
for example, when the heavy quark constituent is the $c$-quark,
\begin{equation} \renewcommand{\arraystretch}{1.3} \begin{array}{l}
\Phi = (D^0, D^+)  \hskip 1cm
\left(\Phi^\dagger = (\bar{D}^0, D^-)^T \right), \\
\Phi^* = (D^{*0}, D^{*+}) \hskip 1cm
\left( \Phi^{*\dagger} = (\bar{D}^{*0}, D^{*-})^T \right).
\end{array} \label{Phi}
\end{equation}
The traditional Lagrangian for the free fields is
\begin{equation}
\textstyle{\cal L}_{\mbox{\scriptsize free}} =
  \partial_\mu \Phi \partial^\mu \Phi^\dagger
  - m_\Phi^2 \Phi \Phi^\dagger
  -\frac12\Phi^{*\mu\nu} \Phi^{*\dagger}_{\mu\nu}
  + m_{\Phi^*}^2 \Phi^{*\mu} \Phi^{*\dagger}_\mu,
\label{Lfree}
\end{equation}
where $\Phi^*_{\mu\nu} \equiv \partial_\mu \Phi^*_\nu - \partial_\nu
\Phi^*_\mu$ is the field strength tensor of the heavy vector meson
fields and $m_\Phi$ and $m_{\Phi^*}$ are the masses of heavy
pseudoscalar and vector mesons, respectively.

In the limit of infinite heavy quark mass, the heavy quark symmetry
implies that the dynamics of the heavy mesons depends trivially on
their spin and mass. Such trivial dependence can be eliminated by
introducing a redefined $4\times4$ matrix field $H(x)$
as\cite{Ge92,Bjorken}
\begin{equation}
H = \frac{1+{v \!\!\!/}}{2} \left( \Phi_v \gamma_5
   - \Phi^*_{v\mu}\gamma^\mu \right),
\label{H}
\end{equation}
instead of the traditional heavy meson fields, $\Phi$ and $\Phi^*$.
Here, we use the conventional Dirac $\gamma$-matrices and
${v \!\!\!/}$ denotes $v_\mu\gamma^\mu$. The fields $\Phi_v$ and
$\Phi_{v\mu}^*$, respectively, represent the heavy pseudoscalar
field and heavy vector fields in the moving frame with a four
velocity $v_\mu$. They are related to the $\Phi$ and $\Phi^*_\mu$
as\cite{Georgi90}
$$\begin{array}{l}
\Phi = e^{-iv\cdot x m_\Phi} \frac{1}{\sqrt{2m_\Phi}} \Phi_v, \\
\Phi^*_\mu = e^{-iv\cdot x m_{\Phi^*}}
\frac{1}{\sqrt{2m_{\Phi^*}}} \Phi^*_{v\mu}.
\end{array}\eqno(\mbox{\ref{H}a})$$
Under the heavy quark spin rotation, $H$ transforms
$$
H \rightarrow S H,
\eqno(\mbox{\ref{H}b})
$$
with $S\in SU(2)_v$ (the heavy quark spin symmetry group boosted by
the velocity $v$). In the heavy meson rest frame, {\it i.e.\,},
$v^\mu=(1,\vec{0})$, $S$ can be written explicitly as
$$
H \rightarrow e^{i\frac12\vec\alpha\cdot\vec\sigma} H,
\eqno(\mbox{\ref{H}c})
$$
where $\sigma_i=\frac12\varepsilon_{ijk}[\gamma_j,\gamma_k]$,
the Dirac spin matrices. To leading order in the heavy meson mass,
the free Lagrangian density describing the heavy mesons propagating
with a four-velocity $v^\mu$ is nothing but that for the freely
propagating heavy quark\cite{Ge92}:
\begin{equation}
{\cal L}_{\mbox{\scriptsize free}}^{HQS}
= - i v_\mu \,\mbox{Tr}(\partial^\mu H \bar{H}).
\label{Lfree2}
\end{equation}
For later convenience, we have introduced $\bar{H} = \gamma_0
H^\dagger \gamma_0$, which transforms under the heavy quark spin
rotation as $\bar{H} \rightarrow \bar{H} S^{-1}$.
One may easily check that it comes as the leading order term in
$m_\Phi(=m_{\Phi^*})$ by substituting Eq.(\ref{H}a) into ${\cal
L}_{\mbox{\scriptsize free}}$.

On the other hand, the dynamics of the light quark system is governed
by the $SU(2)_L\times SU(2)_R$ chiral symmetry, which is realized in
a nonlinear way via a $2\times 2$ unitary matrix
\begin{equation}
\Sigma = \exp\,(\frac{i}{f_\pi}\left(\!\!\begin{array}{cc}
\pi^0 \!&\! \sqrt2\pi^+ \\
\sqrt2\pi^- \!&\! -\pi^0 \end{array}\!\!\right)),
\label{Sigma}
\end{equation}
with the triplet of Goldstone bosons ($\pi^+$, $\pi^0$ and $\pi^-$)
and the pion decay constant $f_\pi$(=93 MeV). Under the
$SU(2)_L\times SU(2)_R$ transformation, $\Sigma$ transforms as
$$
\Sigma \rightarrow L \Sigma R^\dagger, \eqno(\mbox{\ref{Sigma}a})
$$
with global transformations $L\in SU(2)_L$ and $R\in SU(2)_R$.
In terms of $\Sigma$, the interactions among the Goldstone bosons
are described by the Lagrangian density
\begin{equation}
{\cal L}_M = \frac{f_\pi^2}{4}\,\mbox{Tr}(\partial_\mu
\Sigma^\dagger \partial^\mu \Sigma)
+ \cdots ,
\end{equation}
where terms with more derivatives are abbreviated by the ellipsis.

Construction of a chirally invariant Lagrangian for the couplings of
the heavy mesons to the Goldstone bosons can be done by assigning
$H$ a proper transformation rule under the chiral transformation.
There may be a considerable freedom. A standard one is to introduce
a redefined matrix
\begin{equation}
\xi=\sqrt{\Sigma},
\end{equation}
which transforms under the $SU(2)_L\times SU(2)_R$ as
\begin{equation}
\xi \rightarrow L \xi U^\dagger = U \xi R^\dagger.
\label{ct1}
\end{equation}
Here, $U$ is a special unitary matrix depending on $L$, $R$ and the
Goldstone fields. From $\xi$ one can construct a vector field $V_\mu$
and an axial vector field $A_\mu$
\begin{equation}
\renewcommand{\arraystretch}{1.2} \begin{array}{l}
V_\mu = \frac12 (\xi^\dagger \partial_\mu \xi
        + \xi \partial_\mu \xi^\dagger),\\
A_\mu = \frac{i}2 (\xi^\dagger \partial_\mu \xi
        - \xi \partial_\mu \xi^\dagger),
\end{array}
\end{equation}
which have simple chiral transformation properties:
\begin{equation}
\renewcommand{\arraystretch}{1.2} \begin{array}{l}
V_\mu \rightarrow U V_\mu U^\dagger + U \partial_\mu U^\dagger,
\\
A_\mu \rightarrow U A_\mu U^\dagger.
\end{array}
\end{equation}
Let the light quark doublet $q$ transforms as\cite{Georgi84}
\begin{equation}
q=(u,d)^T \rightarrow U q,
\end{equation}
so that the heavy meson field $H(x)$ transforms as
\begin{equation}
H \rightarrow H U^\dagger.
\label{ct2}
\end{equation}
This choice defines a simple transformation rule of $H$ under
the parity operation:
\begin{equation}
\xi({\vec r},t) \rightarrow \xi^\dagger(-{\vec r},t), \hskip 5mm
\mbox{and} \hskip 5mm
H({\vec r},t) \rightarrow \gamma^0 H(-{\vec r},t) \gamma^0.
\end{equation}
In terms of the vector field $V_\mu$, a covariant derivative can be
constructed as
\begin{equation}
\renewcommand{\arraystretch}{1.2} \begin{array}{l}
D_\mu H(x)
= H (\stackrel{\leftarrow}{\partial}_\mu + V^\dagger_\mu),
\end{array}
\end{equation}
which transforms under a chiral transformation as
$D_\mu H \rightarrow (D_\mu H) U^\dagger$.

Now, it is easy to write down a {\em ``heavy-quark-symmetric"}
and {\em ``chirally-invariant"}  Lagrangian for the couplings
of the heavy meson fields to the Goldstone bosons. To the leading
order in the derivatives on the Goldstone boson fields,
it reads
\begin{equation}
{\cal L}_{HQS}={\cal L}_M - i v_\mu \,\mbox{Tr}(D^\mu H \bar{H})
 + g \,\mbox{Tr} (H\gamma^\mu \gamma_5 A_\mu \bar{H}),
\label{LHQS}
\end{equation}
with a universal coupling constant $g$ for the $\Phi\Phi^*\pi$
and $\Phi^*\Phi^*\pi$ interactions. The nonrelativistic quark model
provides a naive estimation\cite{Yan} for the value of $g$ as
\begin{equation}
g=-\textstyle\frac34.
\label{g}
\end{equation}
On the other hand, the Lagrangian leads to the decay widths
\begin{equation}
\Gamma(\Phi^*_{+\frac12}\rightarrow \pi^+\Phi_{-\frac12})
=2\Gamma(\Phi^*_{+\frac12}\rightarrow \pi^0\Phi_{+\frac12})
= \frac{1}{12\pi}\frac{g^2}{f^6_\pi}|\vec{p}_\pi|^3,
\end{equation}
where the subscript $\pm\frac12$ of $\Phi$ and $\Phi^*$ denotes the
third component of their isospin. In case of $Q$=$c$, the $c$-quark,
the experimental upper limit\cite{ACCMOR} of 131 keV on the $D^*$
width implies that $|g|^2\raisebox{-0.6ex}{$\stackrel{\textstyle
<}{\sim}$} 0.5$ when combined with the
$D^{*+}\rightarrow D^+\pi^0$ and $D^{*+}\rightarrow D^0\pi^+$
branching ratios\cite{CLEO}. In this work, however, we will take
the pion decay constant $f_\pi$ and the heavy meson coupling constant
$g$ as adjustable parameters.


\section{Soliton-Heavy Meson Bound State}

With a suitable stabilizing term,\footnote{The explicit form of the
stabilizing terms is not essential for our discussions. Later we will
adopt the Skyrme term as a stabilizing term for numerical results.
For a review of the Skyrme model we refer to Ref.\cite{SK_rev}.}
the nonlinear Lagrangian ${\cal L}_M$ supports a classical soliton
solution\cite{Skyrme}
\begin{equation}
\Sigma_0({\vec r})=\exp(i{\vec\tau}\!\cdot\!{\hat r} F(r)),
\label{SS}
\end{equation}
with the wavefunction $F(r)$ satisfying the boundary conditions
$$
F(0)=\pi \hskip 5mm \mbox{and} \hskip 5mm
F(r)\stackrel{r\rightarrow\infty}{\longrightarrow} 0.
\eqno(\mbox{\ref{SS}a})$$
The soliton solution carries a winding number identified with the
baryon number
$$
\mbox{(B. No.)}=-\frac{1}{24\pi^2}\int\! d^3r
\varepsilon^{ijk}\,\mbox{Tr}(
\Sigma^\dagger_0\partial_i\Sigma_0
\Sigma^\dagger_0\partial_j\Sigma_0
\Sigma^\dagger_0\partial_k\Sigma_0)=1,
\eqno(\mbox{\ref{SS}b})
$$
and a finite mass
$$
M_{sol} = 4\pi \int^\infty_0\!\!\! r^2dr \frac{f_\pi^2}{2}
 \left( F^{\prime 2}+2\frac{\sin^2\!F}{r^2} \right)+\cdots,
\eqno(\mbox{\ref{SS}c})
$$
with $F^\prime=dF/dr$ and the ellipsis denoting the contributions
from the soliton-stabilizing terms.

Our main interest is the bound heavy-meson states (if any) due to the
static potentials provided by the baryon-number-one soliton
configuration (\ref{SS}) sitting at the origin.
Note that we are working with the infinitely heavy soliton.
Explicitly, the potentials are in the form of
\begin{equation}
\renewcommand{\arraystretch}{1.2} \begin{array}{l}
V^\mu=(V^0,\vec{V})=(0,i\upsilon(r){\hat r}\times{\vec\tau}), \\
A^\mu=(A^0,\vec{A})=(0,\frac12(a_1(r){\vec\tau}
      + a_2(r){\hat r}{\vec\tau}\!\cdot\!{\hat r})),
\label{VA}
\end{array}
\end{equation}
with
$$
\upsilon(r)=\frac{\sin^2(F/2)}{r}, \hskip 5mm
a_1(r)=\frac{\sin\!F}{r} \hskip 5mm \mbox{and} \hskip 5mm
a_2(r)=F^\prime - \frac{\sin\!F}{r}.
\eqno(\mbox{\ref{VA}a})
$$

In the rest frame $v^\mu=(1,\vec{0})$, the equation of motion
for the eigenmodes $H_n({\vec r}) e^{-i\varepsilon_n t}$ of the $H$-field
can be read off as
\begin{equation}
\varepsilon_n H_n({\vec r}) = g H_n({\vec r}) \vec{A}\cdot{\vec\sigma} ,
\label{Heq}
\end{equation}
where we have used that
$H(x)\vec{\gamma}\gamma_5 = - H(x)\vec{\sigma}$. Here $\varepsilon_n$ is the
eigenenergy and $n$ denotes a set of quantum numbers which
classify the eigenmodes. The ``hedgehog" configuration (\ref{SS})
correlates the isospin and the angular momentum, while the
heavy-quark symmetry implies the heavy quark spin decoupling from
the set. Thus, the equation of motion is invariant under the parity
operation, the heavy quark spin rotation, and the simultaneous
rotations in the ensemble of spaces: isospin space, ``{\em light-quark}
spin" space, and ordinary space. Let ${\vec L}$, ${\vec S}_\ell$,
${\vec S}_Q$, and ${\vec I}_h$ be the orbital angular momentum,
light quark spin, heavy quark spin, and isospin operators of the
heavy mesons, respectively, and let
$Y_{\ell m}({\hat r})$, $^{}_{\ell}( {\pm\textstyle\frac12 |}$,
${|\pm\textstyle\frac12 )_Q^{}}$, and
$\tilde{\phi}_{\pm\frac12}$ be corresponding eigenstates,
respectively.
(See Appendix~A for their explicit representations.)
The simultaneous rotations mentioned above are generated by the
``{\em light-quark grand spin}" operators defined as
\begin{equation}
{{\vec K}_\ell}={\vec L}+{\vec S}_\ell+{\vec I}_h.
\end{equation}
By the subscript $\ell$, we distinguish ${{\vec K}_\ell}$ from the
traditional grand spin operators
($\vec{K}={\vec L}+{\vec S}+{\vec I}_h$
with ${\vec S}={\vec S}_\ell+{\vec S}_Q$
being the spin operators of the heavy mesons) used in the bound state
approach\cite{CK} in the Skyrme model. Then, the eigenmodes of the
heavy meson can be classified by the third component of heavy quark
spin $s^{}_Q$, the grand spin $(k_\ell,k_3)$ and the parity $\pi$,
so that $n=\{k_\ell,k_3,\pi,s^{}_Q\}$.

The situation is very similar to obtaining the eigenmodes of the
confined quarks in the chiral bag model\cite{BYP}. We start with the
construction of the eigenfunctions of the grand spin and the heavy
quark spin by taking direct products of the four eigenstates,
$Y_{\ell m_\ell}$, $\tilde{\phi}_{\pm\frac12}$,
$^{}_{\ell}( {\pm\textstyle\frac12 |}$, and
${|\pm\textstyle\frac12 )_Q^{}}$, which yields four
${\cal K}^{(i)}_{k_\ell k_3 s^{}_Q}$, $i$=1,2,3,4:
\begin{equation}
{\cal K}^{(i)}_{k_\ell k_3 s^{}_Q} = \displaystyle \sum_{m_s,m_t}
(\ell_i,m_\ell,\textstyle\frac12 ,m_t|\lambda_i,m_\ell+m_t)
(\lambda_i,m_\ell+m_t,\textstyle\frac12 ,m_s|k_\ell,k_3)
Y_{\ell m_\ell}({\hat r}) \tilde{\phi}_{m_t}\,
{^{}_{\ell}( {m_s|} {|s_Q)_Q^{}}},
\end{equation}
with the Clebsch-Gordan coefficients $(\ell_1,m_1,\ell_2,m_2|\ell,m)$.
Here, for a later convenience, we first combine the angular momentum
and the isospin to form $\vec{\lambda}(={\vec L}+{\vec I}_h$) and
then combine the light quark spin with it. Although the heavy quark
spin does not involve in the combination of the light-quark grand
spin, we have included them in the definition of
${\cal K}^{(i)}_{k_\ell k_3 s^{}_Q}$ in order to shorten the
expressions. The explicit forms of
${\cal K}^{(i)}_{k_\ell k_3 s^{}_Q}$ are given in Appendix~B.

In terms of these ${\cal K}^{(i)}_{k_\ell k_3 s^{}_Q}$,
the heavy meson wavefunction can be written as
\begin{equation}
\renewcommand{\arraystretch}{1.5} \begin{array}{ll}
\displaystyle H_n({\vec r}) = \sum_{i=1,2} h^{(i)}_{k_\ell}(r)
{\cal K}^{(i)}_{k_\ell k_3 s^{}_Q} , &
\mbox{for $\pi=-(-1)^{k_\ell}$ states}, \\
\displaystyle H_n({\vec r}) = \sum_{i=3,4} h^{(i)}_{k_\ell}(r)
{\cal K}^{(i)}_{k_\ell k_3 s^{}_Q} , &
\mbox{for $\pi=+(-1)^{k_\ell}$ states},
\end{array}
\end{equation}
with radial functions $h^{(i)}_{k_\ell}(r)$. Note that the electric
modes ($i$=1,2 states) and the magnetic modes ($i$=3,4 states) are
decoupled due to their different parities.

Since the heavy mesons and the soliton are assumed infinitely heavy,
their kinetic effects can be neglected and the heavy mesons are
expected just to sit at the center of the soliton where the potentials
have the lowest value. That is, in the heavy mass limit, all the
radial functions $h^{(i)}_{k_\ell}(r)$ can be approximated as
\begin{equation}
h^i_{k_\ell}(r) = \alpha_i f(r),
\label{hr}
\end{equation}
with a constant $\alpha_i$ and a function $f(r)$ which is strongly
peaked at the origin and normalized as
$ \int^\infty_0 r^2dr |f(r)|^2 = 1 $. The problem is to find the
eigenfunction of the equation
\begin{equation}
\varepsilon {\cal K}_{k^\pi k_3 s_Q} = \textstyle\frac12 gF^\prime(0)
{\cal K}_{k_\ell^\pi k_3 s_Q} \{ ({\vec\tau}\!\cdot\!{\hat r})
[{\vec\sigma}\!\cdot\!{\vec\tau}]({\vec\tau}\!\cdot\!{\hat r})\},
\label{Keq}
\end{equation}
as a linear combination of ${\cal K}^{(i)}_{k_\ell k_3 s_Q}$; {\it
i.e.\,},
${\cal K}_{k_\ell^\pi k_3 s^{}_Q} = \sum_i \alpha_i
{\cal K}^{(i)}_{k_\ell k_3 s_Q}$
($i$=1,2 or 3,4).
In obtaining Eq.(\ref{Keq}), we have used that
$F(r) = \pi + F^\prime(0) r + O(r^3)$ near the
origin so that $a_1(r)\sim -F^\prime(0) + O(r^2)$, and
$a_2(r)\sim 2F^\prime(0) + O(r^2)$ and the identity
$(2{\vec\sigma}\!\cdot\!{\hat r}{\vec\tau}\!\cdot\!{\hat r}
-{\vec\sigma}\!\cdot\!{\vec\tau})=({\vec\tau}\!\cdot\!{\hat r})
({\vec\sigma}\!\cdot\!{\vec\tau})({\vec\tau}\!\cdot\!{\hat r})$.

The expansion coefficients $\alpha_i$ are obtained
by solving the secular equation
\begin{equation}
\sum_{j} {\cal M}_{ij} \alpha_j = - \varepsilon \alpha_i,
\hskip5mm\mbox{ ($i,j$=1,2 or 3,4)}
\label{SecEq}
\end{equation}
where the matrix elements ${\cal M}_{ij}(i,j$=1,2 or 3,4) are defined as
\begin{equation}
{\cal M}_{ij} = \textstyle\frac12gF^\prime(0) \displaystyle
\int\!d\Omega \,\mbox{Tr}\left\{ {\cal K}^{(i)}_{k_\ell k_3 s^{}_Q}
({\vec\tau}\!\cdot\!{\hat r})[({\vec\sigma}\!\cdot\!{\vec\tau})]
({\vec\tau}\!\cdot\!{\hat r})
\bar{{\cal K}}^{(j)}_{k_\ell k_3 s^{}_Q} \right\},
\label{M}
\end{equation}
with
$ \bar{{\cal K}}^{(j)}_{k_\ell k_3 s^{}_Q} = \gamma^0
{\cal K}^{(j)\dagger}_{k_\ell k_3 s^{}_Q} \gamma^0$.
The minus sign in the right hand side of Eq.(\ref{SecEq}) comes
from the fact that the basis states ${\cal K}^{(i)}_{k_\ell k_3
s^{}_Q}$ are normalized as (\ref{Knorm}).

Explicit matrix elements are presented in Appendix~B, according
to which ${\cal K}^{(1,2)}_{k_\ell k_3 s^{}_Q}({\hat r})$ are
already the eigenstates of Eq.(\ref{Keq}) with the degenerate
eigenenergy $\frac12gF^\prime(0)$ and ${\cal M}$ should be
diagonalized for $i,j=3,4$. The diagonalization of
${\cal M}_{ij}(i,j$=3,4) leads to two eigenstates:
\begin{equation}
\renewcommand{\arraystretch}{1.7}\begin{array}{ll}
{\cal K}^{(+)}_{k_\ell k_3 s_Q} =
\sqrt{\frac{k_\ell}{2k_\ell+1}} {\cal K}^{(3)}_{k_\ell k_3 s_Q}
+ \sqrt{\frac{k_\ell+1}{2k_\ell+1}} {\cal K}^{(4)}_{k_\ell k_3 s_Q},
& (\varepsilon=+\textstyle\frac12gF^\prime(0)) \\
{\cal K}^{(-)}_{k_\ell k_3 s_Q} =
\sqrt{\frac{k_\ell+1}{2k_\ell+1}} {\cal K}^{(3)}_{k_\ell k_3 s_Q}
 - \sqrt{\frac{k_\ell}{2k_\ell+1}} {\cal K}^{(4)}_{k_\ell k_3 s_Q}.
& (\varepsilon=-\textstyle\frac32gF^\prime(0))
\end{array}
\end{equation}
Thus, for each set of different quantum numbers
$\{ k_\ell(\neq 0), k_3, s^{}_Q\}$,
\footnote{In case of 
$k_\ell=0$, we have two eigenstates;
$$\renewcommand{\arraystretch}{1.3}\begin{array}{ll}
{\cal K}^{(1)}_{0 0 s_Q}({\hat r}): &
 \varepsilon=+\frac12gF^\prime(0), \\
{\cal K}^{(3)}_{0 0 s_Q}({\hat r}): &
 \varepsilon=-\frac32gF^\prime(0).
\end{array}$$
} 
we have one state with the eigenenergy
$\varepsilon=-\frac32gF^\prime(0)$
and three degenerate states with $\varepsilon=\frac12gF^\prime(0)$.
Since $g<0$ and $F^\prime(0)<0$ (in case of baryon-number-one
soliton solution), we have one {\em soliton-heavy meson bound state}
with a binding energy $\frac32gF^\prime(0)$.
The positive energy states imply three (one for $k_\ell=0$)
{\em soliton-antiflavored heavy meson bound states} with a
binding energy $\frac12gF^\prime(0)$.\cite{OPM0} (See Sec.~VI)
In Fig.~1 given are energy levels of the bound states
of heavy mesons and antiflavored ones. Shaded areas denote the
continuum states and triply degenerate states are represented as
thick lines.

In case of a typical soliton solution stabilized by the Skyrme
term\cite{Skyrme}, we have $F^\prime(0)\sim -2ef_\pi$ with $e$
being the Skyrme parameter. When the parameters are fixed as
$f_\pi$=64.5 MeV and $e$=5.45 for the soliton to fit well the
nucleon and Delta masses\cite{ANW}, $F^\prime(0)$ amounts to
$\sim -0.70$ GeV which implies that the binding energies of
the soliton-heavy meson and soliton-antiflavored heavy meson
bound states are $\frac32gF^\prime(0)\sim 0.79$ GeV and
$\frac12gF^\prime(0)\sim 0.26$ GeV, respectively. Compared with
those of Refs.\cite{RRS2,RS93}, one can see that the binding
energies are reduced by a factor half and more. It should be
emphasized further that in Refs.\cite{RRS2,RS93} the binding
energy increases as the heavy meson mass increases and also
that our results are obtained with infinite heavy meson mass.

Each state can be combined with the heavy quark spin to produce
doubly-degenerate grand spin eigenstates with $k_\pm = k_\ell \pm 1/2$
(provided $k_\ell\neq 0$, for which we have only a grand spin
$k=1/2$ state). These degeneracies are the direct consequence of the
heavy quark symmetry and a special attention should be paid to the
quantization procedure, although the degeneracy in $k_\ell$ (and in
principal quantum numbers which are related with the radial
excitations) may be an artifact from the approximation
(\ref{hr}) on the radial function $h^{(i)}_{k_\ell}(r)$.
In general, when the heavy meson's kinetic term is taken into account,
the radial function feels the centrifugal potential
$\ell_{{\mbox{\scriptsize eff}}}(\ell_{\mbox{\scriptsize eff}}+1)/r^2$
near the origin so that it behaves as
$h^{(i)}_{k_\ell}\sim r^{\ell_{\mbox{\scriptsize eff}}}$.
Here, $\ell_{\mbox{\scriptsize eff}}$ is the ``effective" angular
momentum\cite{CK}, which is related to $\ell$ as
\begin{equation}
\ell_{\mbox{\scriptsize eff}} = \left\{
\renewcommand{\arraystretch}{1.2}\begin{array}{ll}
 \ell+1, & \mbox{if $\lambda=\ell+\frac12$}, \\
 \ell-1, & \mbox{if $\lambda=\ell-\frac12$}.
\end{array} \right.
\end{equation}
Due to the vector potential
$\vec{V}$ ($\sim i({\hat r}\times\vec\tau)/r$, near the origin)
the singular structure of $\vec{D}^2=(\vec{\nabla}-\vec{V})^2$
is altered from $\ell(\ell+1)/r^2$ (that of $\vec\nabla^2$) to
$\ell_{\mbox{\scriptsize eff}}(\ell_{\mbox{\scriptsize eff}}+1)/r^2$.
Thus, only those states with $\ell_{\mbox{\scriptsize eff}}=0$ can
have strongly peaked radial function and the degeneracies will be
broken in such a way that the states with higher
$\ell_{\mbox{\scriptsize eff}}$ have the higher energy.
For the positive parity states,
$\ell_{\mbox{\scriptsize eff}}=0$ can be achieved only when $\ell=1$.
In Table~I listed are a few heavy meson eigenstates which involve
$\ell$=0,1 angular basis. From now on, we will restrict our
considerations to these states. Note that the
$\ell_{\mbox{\scriptsize eff}}$ of the wavefunctions
${\cal K}^{(1)}_{1ms^{}_Q}$,
${\cal K}^{(4)}_{2ms^{}_Q}$
(included in ${\cal K}^{(\pm)}_{2ms^{}_Q}$), and
${\cal K}^{(3)}_{2ms^{}_Q}$
(included in ${\cal K}^{(\pm)}_{2ms^{}_Q}$) is 2 and
that the ${\cal K}^{(4)}_{1ms^{}_Q}$
(included in ${\cal K}^{(\pm)}_{1ms^{}_Q}$)
is 1. In case of the finite heavy meson mass, these states will
have higher eigenenergies and will become even unbound.

In terms of the eigenmodes, we can expand the heavy meson field
operator as
\begin{equation}
H(x) = \sum_{n} H_n({\vec r}) e^{-i\varepsilon_n t} a_n,
\label{Hop}
\end{equation}
with the heavy meson annihilation operators $a_n$. Note that we
don't need to include the term for the antiparticles. The Fock
states on which the quark field operators act are obtained as
\begin{equation}
|n_1,n_2,\cdots\rangle=a_{n_1}^\dagger a_{n_2}^\dagger \cdots
|\mbox{vac}\rangle,
\end{equation}
where $|\mbox{vac}\rangle$ is the vacuum of the heavy meson fields.
Hereafter, we will denote the Fock states of a single heavy
meson occupying the corresponding state as in Table I. (To simplify
the notations, unless necessary, we will not specify such
trivial quantum numbers as the third component of the grand spin,
the heavy quark spin and the parity; $k_3$, $s_Q$ and $\pi$.)


\section{Collective Coordinate Quantization}

What we have obtained so far is the soliton-heavy meson (or antiflavored
heavy meson) bound state which carries a baryon number and a heavy
flavor (or anti-heavy flavor).
To endow the states with correct quantum numbers such as spin and
isospin, we have to go to next order in $1/N_c$,
while remaining in the same order in $m_Q$, namely $O(m_Q^0 N_c^{-1})$.
This can be done by quantizing the zero modes associated with the
invariance under simultaneous $SU(2)$ rotation of the soliton
configuration together with the heavy meson fields:
\begin{equation}
\xi^{}_0 \rightarrow C \xi^{}_0 C^\dagger, \hskip 5mm \mbox{and}
\hskip 5mm
H \rightarrow H C^\dagger,
\end{equation}
with an arbitrary constant $SU(2)$ matrix $C$ and $\xi^2_0
\equiv U_0$. The rotation becomes dynamical by giving time
dependence to the $SU(2)$ collective variables as
\begin{equation}
\xi({\vec r},t) = C(t) \xi^{}_0({\vec r}) C^\dagger(t), \hskip 5mm
\mbox{and} \hskip 5mm
H({\vec r},t) = H_{{\mbox{\scriptsize bf}}}({\vec r},t) C^\dagger(t),
\label{CV}
\end{equation}
and then the quantization is done by elevating the collective variables
to the corresponding quantum mechanical operators.
In Eq.(\ref{CV}), $H_{{\mbox{\scriptsize bf}}}$ refers to the heavy
meson field in the isospin-co-moving frame, while $H({\vec r},t)$
refers to that in the laboratory frame. Substitution of Eq.(\ref{CV})
into Eq.(\ref{LHQS}) leads us to the Lagrangian (in the reference
frame where the heavy meson is at rest in space but rotating in
isospin space)
\begin{equation}
\renewcommand{\arraystretch}{1.5} \begin{array}{l}
\textstyle L^{{\mbox{\scriptsize rot}}}
= - M_{sol}
  + \displaystyle\int\!d^3r \left\{ -i\,\mbox{Tr}(\partial_0
H_{\mbox{\scriptsize bf}} \bar{H}_{\mbox{\scriptsize bf}})
  + g\,\mbox{Tr}(H_{\mbox{\scriptsize bf}}\,
{\vec A}\!\cdot\!{\vec\sigma} \, \bar{H}_{\mbox{\scriptsize bf}} )
\right\} \\
\hskip 1cm
+ \textstyle\frac12 {\cal I}\omega^2
  - \textstyle\frac12 \displaystyle\int\!d^3r \left\{ \,\mbox{Tr}(
H_{\mbox{\scriptsize bf}}
\textstyle \frac12(\xi^\dagger \vec \tau \cdot \vec \omega \xi + \xi \vec
\tau \cdot \vec \omega \xi^\dagger)\,
\bar{H}_{\mbox{\scriptsize bf}} ) \right\},
\end{array}\label{Lcol}
\end{equation}
where we have kept terms up to $O(m^0_QN_c^{-1})$.
The ``angular velocity", ${\vec\omega}$,  of the collective
rotation is defined by
$$
C^\dagger\partial_0 C \equiv \textstyle\frac12
i{\vec\tau}\!\cdot\!{\vec\omega}, \eqno(\mbox{\ref{Lcol}a})
$$
and ${\cal I}$ is the moment of inertia of the soliton configuration
with respect to the rotation
$$
{\cal I} = \textstyle\frac{8\pi}{3}f^2_\pi\!\displaystyle
           \int^\infty_0\!\!\!r^2dr (\sin^2 F+\cdots)
\eqno(\mbox{\ref{Lcol}b})$$
where the contributions from the soliton-stabilizing-Lagrangian
are abbreviated simply by ellipsis.

Given the Lagrangian (\ref{Lcol}) that describes dynamics up to order
$O(m_Q^0N_c^{-1})$, one has equation of motion
\begin{equation}
\textstyle i\partial_0 H_{{\mbox{\scriptsize bf}}}
= H_{{\mbox{\scriptsize bf}}}\,[ g {\vec A}\!\cdot\!{\vec\sigma}
- \textstyle\frac14 (\xi^\dagger \vec\tau \cdot \vec\omega \xi + \xi \vec\tau
\cdot \vec \omega \xi^\dagger) ]
\end{equation}
consistent to that order. The last ``Coriolis" term in the equation of
motion couples the fast and slow degrees of freedom.
Although the heavy mesons are infinitely heavy, their angular
momentum and isospin are associated with the light constituents. Thus,
we may take those light degrees of freedom of the heavy meson fields
as the fast ones and the collective rotations as the slow ones.
Note that the scale of the eigenenergies $|\varepsilon_n|$
of the heavy mesons is much greater than that of the
rotational velocity; $|\varepsilon_n|\gg |\omega|$.

Generally accepted procedure of handling these different scale is
as follows.\cite{LNRZ} We first solve the equation of motion for
fast degrees of freedom with slow degrees of freedom ``frozen". In
this way, we get ``snap-shot" pictures of the fast motion. Next we
solve the equation of motion for slow degrees of freedom taking into
account the ``relic" of the fast motion that has been ``integrated
out", in a manner completely analogous to the incorporation of Berry
phases\cite{SW}.  It is also analogous to the ``strong-coupling
limit" of the particle-rotor model\cite{BM} in nuclear physics, where
the coupling between the rotating ``core" and the particle is much
stronger than the perturbation of the single-particle motion by
Coriolis interaction. Here, the roles of the particle and the rotor
are played by the bound heavy-mesons and the rotating soliton
configuration. Thus, we may take the assumption that the bound heavy
mesons rotate together with the soliton core in the {\em unchanged
eigenmodes}. It enables us to expand the
$H_{{\mbox{\scriptsize bf}}}(x)$ in terms of
the classical eigenmodes obtained in Sec.~III as Eq.(\ref{Hop}):
\begin{equation}
H_{{\mbox{\scriptsize bf}}}(x)
=\sum_n H_n^{}({\vec r})e^{-i\varepsilon_n^{}t} a_n^{}.
\end{equation}

Taking Legendre transformation of the Lagrangian,
we obtain the Hamiltonian as
\begin{equation}
\renewcommand{\arraystretch}{1.4} \begin{array}{rl}
H \!\!\! &= \displaystyle \int\!d^3r \left\{
\frac{\delta{\cal L}^{{\mbox{\scriptsize
rot}}}}{\delta(\dot{H}_{{\mbox{\scriptsize bf}},\alpha\beta})}
\dot{H}_{{\mbox{\scriptsize bf}},\alpha\beta} \right\}
+\frac{\delta L^{{\mbox{\scriptsize rot}}}}{\delta\omega_a}\omega_a
- L^{\mbox{\scriptsize rot}} \\
 & \displaystyle = M_{sol} - g\!\int\!d^3r\,
\,\mbox{Tr}(H_{{\mbox{\scriptsize bf}}}\,{\vec A}\!\cdot\!{\vec\sigma}
\,\bar{H}_{{\mbox{\scriptsize bf}}})
+ \frac{1}{2{\cal I}}({\vec R}-{\vec\Theta}(\infty))^2 ,
\end{array} \label{Hcol}
\end{equation}
where the rotor spin ${\vec R}$ is the canonical momenta conjugate
to the collective variables $C(t)$:
$$
R_a \equiv \frac{\delta L^{{\mbox{\scriptsize rot}}}}{\delta\omega^a}
= {\cal I} \omega_a + \Theta_a(\infty),
\eqno(\mbox{\ref{Hcol}a})
$$
with ${\vec\Theta}(\infty)$ defined as
$$
{\vec\Theta}(\infty) \equiv  - \textstyle\frac12
\displaystyle \int\! d^3r\,\mbox{Tr} [ H_{\mbox{\scriptsize
bf}} \,
\textstyle \frac12 (\xi^\dagger \vec \tau \xi + \xi \vec \tau \xi^\dagger )
\, {\bar H}_{\mbox{\scriptsize bf}} ],
\eqno(\mbox{\ref{Hcol}b})
$$
whose expectation value with respect to the state $|n\rangle$ is the
Berry phase associated with the collective rotation\cite{NRZ,LNRZ}.
Note that {\em it is nothing but the isospin of the heavy mesons modulo
the sign}. (See Eq.(\ref{I}b) below)

With the collective variable introduced in Eq.(\ref{CV}), the
isospin of the fields $U(x)$ and $H(x)$ is entirely shifted to
$C(t)$. To see this, consider the isospin rotation
\begin{equation}
   \Sigma \rightarrow {\cal A} \Sigma {\cal A}^\dagger, \hskip 8mm
   H \rightarrow H {\cal A}^\dagger,
\end{equation}
with ${\cal A} \in SU(2)_V$, under which the collective variables
and fields in body-fixed frame transform as
\begin{equation}
C(t) \rightarrow {\cal A} C(t), \hskip 8mm
H_{{\mbox{\scriptsize bf}}}(x)
\rightarrow H_{{\mbox{\scriptsize bf}}}(x).
\end{equation}
Note that the $H$-field becomes
isospin blind in the (isospin) co-moving frame. The conventional
Noether construction leads the isospin of the system as
\begin{equation}
I_a = \textstyle\frac12 \,\mbox{Tr}(\tau_a C \tau_b C^\dagger)
( {\cal I} \omega_b + \Theta_b(\infty) ) =
D^{\mbox{\scriptsize}}_{ab}(C) R_b,
\label{HB-iso}
\end{equation}
with $D^{\mbox{\scriptsize}}_{ab}(C)$ being the $SU(2)$ adjoint
representation associated with the collective variables $C(t)$.

Under a spatial rotation (together with the spin rotation in
case of the heavy meson), with the help of the $K$-symmetry,
the fields transform as
\begin{equation} \renewcommand{\arraystretch}{1.4} \begin{array}{l}
\Sigma({\vec r},t) \rightarrow   \Sigma(\vec{r}',t)
= C(t) {\cal B}^\dagger \Sigma_0({\vec r}) {\cal B} C^\dagger(t), \\
H({\vec r},t) \rightarrow
e^{i\frac12\vec{\alpha}\cdot{\vec\sigma}} H(\vec{r}',t)
e^{-i\frac12\vec{\alpha}\cdot{\vec\sigma}}
= ( e^{i\frac12\vec{\alpha}\cdot{\vec\sigma}}
H_{{\mbox{\scriptsize bf}}}(\vec{r}',t)
e^{-i\frac12\vec{\alpha}\cdot({\vec\sigma}+{\vec\tau})})
 {\cal B}^\dagger C(t),
\end{array}
\end{equation}
with $\vec{r}'=\exp(i\vec\alpha\cdot{\vec L}){\vec r}$ and
${\cal B}=\exp(i\frac12\vec{\alpha}\cdot{\vec\tau})\in SU(2)$.
This means that the spatial rotation is equivalent to the
transformation of the collective variables and $H$-fields in the
body fixed frame as
\begin{equation}
\renewcommand{\arraystretch}{1.4} \begin{array}{l}
C(t) \rightarrow C(t) {\cal B}^\dagger,  \\
H_{{\mbox{\scriptsize bf}}}({\vec r},t) \rightarrow
e^{i\frac12\vec{\alpha}\cdot{\vec\sigma}}
H_{{\mbox{\scriptsize bf}}}(\vec{r}',t)
e^{-i\frac12\vec{\alpha}\cdot({\vec\sigma}+{\vec\tau})}.
\end{array}
\end{equation}
Therefore, we get the fact that the spin of the $H_{{\mbox{\scriptsize
bf}}}(x)$ is the
grand spin; that is, the isospin of the $H$-field is transmuted
into the part of the spin in the isospin co-moving frame.
Remember that the $H_{{\mbox{\scriptsize bf}}}(x)$ becomes isospin
blind in that frame.
Applying the Noether theorem to the Lagrangian (\ref{Lcol}),
we obtain the spin of the system explicitly as
\begin{equation}
{\vec J}={\vec R} + \vec{K}_{{\mbox{\scriptsize bf}}},
\label{Jbf}
\end{equation}
with the grand spin of the heavy meson fields
in the isospin co-moving frame
\begin{equation}
 \vec{K}_{{\mbox{\scriptsize bf}}} = - \int\!\!d^3r\;
\textstyle \,\mbox{Tr}\{({\vec L} H_{{\mbox{\scriptsize bf}}} +
[\textstyle\frac12 {\vec\sigma},H_{{\mbox{\scriptsize bf}}}]
  + H_{{\mbox{\scriptsize bf}}}(-\textstyle\frac12
{\vec\tau})){\bar H}_{{\mbox{\scriptsize bf}}}\}.
\end{equation}

Finally, the heavy-quark spin symmetry of the Lagrangian under the
transformation
\begin{equation}
H(x) \rightarrow e^{i\frac12\vec{\alpha}\cdot{\vec\sigma}}
H(x) = (e^{i\frac12\vec{\alpha}\cdot{\vec\sigma}}
       H_{{\mbox{\scriptsize bf}}}(x)) C(t),
\end{equation}
has nothing to do with the collective rotations. The heavy-quark spin
operator remains unchanged in the isospin co-moving frame :
\begin{equation}
\vec S_Q =-\int\!\!d^3r\;\,\mbox{Tr}
(\textstyle\frac12 {\vec\sigma} H{\bar H})
 = \displaystyle -\int\!\!d^3r\;\,\mbox{Tr}
(\textstyle\frac12 {\vec\sigma} H_{{\mbox{\scriptsize bf}}}{\bar
H}_{\mbox{\scriptsize bf}}).
\end{equation}
Because of this heavy-quark spin decoupling, it is convenient to
proceed with the spin operators ${\vec J}_\ell$ for the light
degrees of freedom in the soliton-heavy meson bound system
defined as
\begin{equation}
{\vec J}_\ell = {\vec J} - {\vec S}_Q = {\vec R} + {{\vec K}_\ell}.
\label{Jell}
\end{equation}

Upon canonical quantization, the collective variables become the
quantum mechanical operators; the isospin (${\vec I}$), the spin
(${\vec J}_\ell$) and the  spin of the rotor (${\vec R}$) discussed
so far become the corresponding operators ${\tilde I}_a$,
${\tilde J}_{\ell,a}$ and ${\tilde R}_a$, respectively.
We distinguish those operators associated with the collective
coordinate quantization by using a tilde on them. Let the eigenstates
of the rotor-spin operator ${\tilde R}_a$ be denoted by
$|i;m_1,m_2\}$ $(m_1,m_2=-i,-i+1, \cdots, i)$:
\begin{equation}
\renewcommand{\arraystretch}{1.2} \begin{array}{c}
{\tilde R}^2 |i;m_1,m_2\} = i(i+1)|i;m_1,m_2\}, \\
{\tilde R}_z |i;m_1,m_2\} = m_2 |i;m_1,m_2\}, \\
{\tilde I}_z |i;m_1,m_2\} = m_1 |i;m_1,m_2\}.
\end{array}
\end{equation}
These states are represented explicitly by the Wigner $D$-functions
\begin{equation}
\sqrt{2i+1} D^{(i)}_{m_1,m_2}(C).
\label{WigD}
\end{equation}

At this point, it should be mentioned that we are following exactly
the same quantization procedure of CK\cite{CK}. We quantize the whole
soliton-heavy meson bound system to obtain the heavy baryon states. In
Refs.\cite{JMW,HB}, only the soliton is quantized to nucleons and
$\Delta$'s by using the collective coordinate quantization. Then the
heavy mesons with good isospin and spin are bound to form a heavy
baryon. The corresponding isospin and spin operator of the heavy
baryon system are different from Eqs. (\ref{HB-iso}) and (\ref{Jbf}),
respectively. However, both approaches lead to the same final
results for the heavy baryon spectrum.


\section{Heavy Baryons}

The eigenstates $|i,i_3;j_\ell,j_{\ell,3};s^{}_Q\rangle\!\rangle$
of the operators ${\tilde I}_a$ and ${\tilde J}_{\ell,a}$
with their corresponding quantum numbers $i,i_3$ (isospin) and
$j_\ell,j_{\ell,3}$ (spin of the light degrees of freedom)
are given by the linear combinations of the direct
product of the the rotor-spin eigenstates $|i;m_1,m_2\}$
and the single-particle Fock state $|n\rangle$:
\begin{equation}
|i,i_3;j_\ell,j_{\ell,3};s^{}_Q\rangle\!\rangle_{a}^{}
=\sum_{m} (i,j_{\ell,3}-m,k_\ell^a,m|j_\ell,j^{}_{\ell,3})
|i;i_3,j^{}_{\ell,3}-m\} |k_\ell,m,s^{}_Q\rangle_a^{}.
\label{BS}
\end{equation}
One may combine further the heavy quark spin and the spin of the
light degrees of freedom to construct the states with a good total spin,
which is, however, not necessary in our discussion.
Remember that, in the infinite heavy quark mass limit,
$(j_\ell,j_{\ell,3})$ themselves are good quantum numbers to
label the heavy hadrons. For a given set of $(i,j_\ell)$,
there can be more than one state depending on which Fock state
$|n\rangle$ is involved in the combination (\ref{BS}). We will
distinguish them by using a sequential number, $a$(=1,2$\cdots$),
in $|i,j_\ell\rangle\!\rangle_1$. Here again, to shorten the
expressions, we will not specify the quantum numbers
$i_3$, $j_3$ and $s^{}_Q$ unless necessary.

In Table II, we list a few $|i,j_\ell^\pi\rangle\!\rangle$
states for low-lying heavy baryons. Here, we consider the
$|i,j_\ell^\pi\rangle\!\rangle$ states made of the {\em integer}
rotor-spin-states so that they describe the heavy baryons of
a half-integer spin ($j=j_\ell\pm\textstyle\frac12 $). In the
Table, we do not present the states such as
$|0,1^+\rangle\!\rangle$, $|1,0^+\rangle\!\rangle$ and
$|0,0^-\rangle\!\rangle$, which cannot form the bound states.
The incorporation of the $1/N_c$ order corrections due to collective
rotations cannot turn them into bound systems.
We also exclude two other possible $|1,1^-\rangle\!\rangle$ states
from the list, since they come from the
heavy-meson states $|2\rangle_{1,2}$ of $\ell=2$ and thus they do not
mix up with the state $|1,1^-\rangle\!\rangle$ shown in Table II
by the collective rotation.

The physical heavy baryons of our concern appear as the eigenstates
of the Hamiltonian ${\tilde H}$. We will not try to find the exact
eigenstates of the Hamiltonian. Remind that we have kept the terms
only up to $O(m_Q^0 N_c^{-1})$ in the collective coordinate
quantization procedure. We take the last term in the Hamiltonian
\begin{equation}
{\tilde H}_{rot} \equiv \frac{1}{2{\cal I}}
({\vec R} -{\vec\Theta}(\infty))^2,
\end{equation}
as a perturbation of order $O(m_Q^0 N_c^{-1})$ and we will
consistently search for the approximate eigenstates to that order.

Except for the case of $|1,1^+\rangle\!\rangle$, we have only one
bound state for a given $(i,j^\pi_\ell)$. To the first order,
the eigenstate of ${\tilde H}$ is approximated by the unperturbed
one $|i,j_\ell^\pi\rangle\!\rangle_1$ and the mass correction
to the corresponding baryon is obtained by taking the
expectation value of the ${\tilde H}_{rot}$ with respect to it:
\begin{equation}
M_{(i,j^\pi_\ell)}^{rot}= \frac{1}{2{\cal I}}
\langle\!\langle i,j^\pi_\ell |({\vec R}
- {\vec\Theta}(\infty))^2 | i,j^\pi_\ell \rangle\!\rangle.
\end{equation}
Now, our problem is reduced to evaluating the expectation values
of the operators $-2{\vec R}\cdot{\vec\Theta}(\infty)$
and $\vec\Theta^2(\infty)$. Using Eq.(\ref{WE1}), we easily obtain
the expectation value of the operator
$-2{\vec R}\cdot{\vec\Theta}(\infty)$ as
\begin{equation}
{}_a\langle\!\langle i,j_\ell^\pi
|-2{\vec R}\cdot{\vec\Theta}(\infty) |i,j_\ell^\pi\rangle\!\rangle_a
= c_{aa}^{} \{ j_\ell(j_\ell+1)-i(i+1)-k_\ell(k_\ell+1) \},
\end{equation}
with $c_{aa}^{}$ being the $c$-value associated with the
single-particle Fock state $|k_\ell\rangle_a$ participating
in the construction of the state $|i,j_\ell^\pi\rangle\!\rangle$.
As for the expectation value of the operator $\vec\Theta^2(\infty)$,
in many CK-based models\cite{RRS2} it has been approximated as
\begin{equation}
\langle\!\langle i,j_\ell^\pi |\vec\Theta^2(\infty)
|i,j_\ell^\pi\rangle\!\rangle
={^{}_a\langle k_\ell| \vec\Theta^2(\infty) |k_\ell\rangle_a^{}}
\approx |{^{}_a\langle k_\ell| {\vec\Theta}(\infty)
|k_\ell\rangle_a^{}}|^2 = c_{aa}^2 k_\ell(k_\ell+1).
\label{Aprx}
\end{equation}
In the heavy meson mass limit, it can be exactly evaluated as
\begin{equation}
{^{}_a\langle k_\ell,k_3|\vec\Theta^2(\infty)
|k_\ell,k_3\rangle_a^{}} = \sum_{\{k_\ell^\prime,k_3^\prime,b\}}
|{^{}_a\langle k_\ell,k_3|{\vec\Theta}(\infty)|k'_\ell,k'_3
\rangle_b^{}}|^2,
\end{equation}
where the summation runs over the complete set of intermediate Fock
states $|k'_\ell,k'_3\rangle_b^{}$. Including all the Fock states
that have non-vanishing expectation value
$\langle n|{\vec\Theta}(\infty)|m \rangle$ with the help of the
approximation (\ref{hr}) on the radial functions, we obtain
\begin{equation}
{^{}_a\langle k_\ell|\vec\Theta^2(\infty)|k_\ell\rangle_a^{}}
= \textstyle\frac34.
\label{Exct}
\end{equation}
One may obtain the same result by using the fact that
${\vec\Theta}(\infty)$ defined as in Eq.(\ref{Hcol}b) is nothing
but the isospin operator ${\vec I}_h$ (modulo opposite sign) of
the heavy meson field.

Thus, for the heavy baryons with quantum numbers $(i,j_\ell^\pi)$
which allow only one bound $|i,j_\ell^\pi\rangle\!\rangle$ state,
we obtain the mass formula as
\begin{equation}
m_{(i,j_\ell^\pi)} = M_{sol} + \omega_B^{} + \frac{3}{8{\cal I}}
+ \frac{1}{2{\cal I}} \{ cj_\ell(j_\ell+1) + (1-c)i(i+1)
- c k_\ell(k_\ell+1) \},
\label{MF1}
\end{equation}
where $c$ is an abbreviation for $c_{aa}$
and $\omega_B \equiv \overline{m}_\Phi - \frac32gF^\prime(0)$.
In order to compare it with Eq.(\ref{MF}), we have included
the weight averaged heavy-meson mass
$\overline{m}_{\Phi}(\equiv\frac14(3m_{\Phi^*}^{}+m_{\Phi}^{}))$.
Eq.(\ref{MF1}) is in a quite different form from Eq.(\ref{MF})
and satisfies the heavy-quark symmetry regardless of the
$c$-value. However, all the $c$-values associated with the heavy-meson
bound states $|k_\ell\rangle_-$ vanish identically; {\it viz.},
\begin{equation}
c_{--}^{}
= \frac{k_\ell+1}{2k_\ell+1} c^{}_{33}
  - 2\frac{\sqrt{k_\ell(k_\ell+1)}}{2k_\ell+1} c^{}_{34}
  + \frac{k_\ell}{2k_\ell+1} c_{44}^{}
=0 ,
\end{equation}
as given in Appendix~B. Consequently, both mass formulas (\ref{MF})
and (\ref{MF1}) yield the same heavy baryon masses apart from a
constant shift coming from the way of evaluating the expectation
value of $\vec\Theta^2(\infty)$ as in (\ref{Aprx}) and (\ref{Exct}).
It should be emphasized that such a coincidence owes entirely to
the vanishing $c$-values and the presence of the unique bound state.
When one has non-vanishing $c$-value as we shall see in Sec.~VI,
Eq.(\ref{MF}) cannot be applied anymore.

If we have multiple degenerate bound states,
$|i,j_\ell^\pi\rangle\!\rangle_a(a=1,2,\cdots)$, the situation
becomes a little bit complicated. The mass corrections and the
corresponding eigenstate are obtained by diagonalizing
the energy matrix ${\cal E}$ whose matrix element is defined by
\begin{equation}
{\cal E}_{ab} = {^{}_a\langle\!\langle i,j^\pi_\ell |
{\tilde H}_{rot} | i,j^\pi_\ell \rangle\!\rangle_b^{}}.
\hskip5mm(a,b=1,2,\cdots)
\end{equation}
For the $(i=1,j_\ell^\pi=1^+)$ heavy baryons, we have doubly
degenerate bound states; $|1,1\rangle\!\rangle_1$ and
$|1,1\rangle\!\rangle_2$. Since they are, respectively, made of
$|k_\ell=0\rangle_3$ and $|k_\ell=2\rangle_-$, the first rank tensor
${\vec\Theta}(\infty)$ cannot lead to nonvanishing energy matrix
between the two states. Thus, each state can be separately the
eigenstate with
their degenerate mass\footnote{The  
exact degeneracy is however an artifact of the approximation
of using the same radial function for all the state. When the
heavy mesons are allowed to move, $|1,1\rangle\!\rangle_2$ will
have higher energy.} 
given by the mass formula (\ref{MF1}).

Because of vanishing $c$-value, we can write the mass formula for
the heavy baryon simply as
\begin{equation}
m_{(i,j_\ell^\pi)}^{} = M_{sol} + \omega_B +
\frac{1}{2{\cal I}}(i(i+1)+\textstyle\frac34).
\end{equation}
In Fig.~2, the resulting heavy baryon spectrum is presented
schematically. It is interesting to observe the degeneracy in mass
of the heavy baryons with positive and negative parities.
Since we are working upon an assumption on the radial functions
with ignoring any effects of the kinetic term,  we are not at a
position to conclude whether such a parity doubling has any
physical importance or it is just an artifact of the approximation.

Explicitly, we have the masses of $(0,0^+)$ and $(1,1^+)$ heavy
baryons as
\begin{equation}
\renewcommand{\arraystretch}{1.5} \begin{array}{l}
m_{(0,0^+)} = M_{sol} + \omega_B^{} + 3/8{\cal I}, \\
m_{(1,1^+)} = M_{sol} + \omega_B^{} + 11/8{\cal I}, \\
\end{array} \label{MF2}
\end{equation}
which correspond to the mass of $\Lambda_Q$ and degenerate mass
of $\Sigma_Q$ and $\Sigma_Q^*$, respectively and are consistent
with the results of Ref.\cite{JMW}. Eq.(\ref{MF2}) yields an
interesting model-independent relation for the mass difference
of $\Sigma_Q(\Sigma^*_Q)$ and $\Lambda_Q$:
\begin{equation}
m_{\Sigma_Q,\Sigma^*_Q}^{} - m_{\Lambda_Q}^{} = \textstyle\frac23
( m_\Delta - m_N ) \approx 0.20 \mbox{ GeV},
\end{equation}
where we have used the fact that the $SU(2)$ collective quantization
of the bare soliton leads to the the nucleon and delta masses as
\begin{equation}
m^{}_N = M_{sol} + \frac{3}{8{\cal I}}, \hskip 5mm \mbox{and}
\hskip 5mm
m^{}_\Delta = M_{sol} + \frac{15}{8{\cal I}}.
\end{equation}

In the mass formula for the heavy baryons, we have three
parameters, $M_{sol}$, $1/{\cal I}$ and $gF^\prime(0)$. For a
naive prediction on the heavy baryon masses, we fit them to
produce experimental values of
$m_N^{}$(=939 MeV), $m_\Delta^{}$(=1232 MeV) and
$m_{\Lambda_c}^{}$(=2285 MeV), which leads to
\begin{equation}
M_{sol}= 866 \mbox{ MeV}, \hskip 5mm
1/{\cal I} = 195 \mbox{ MeV} \hskip 5mm \mbox{and} \hskip 5mm
gF^\prime(0) = 419 \mbox{ MeV}.
\label{PMTS}
\end{equation}
Combined with the slope of the soliton wavefunction
$F^\prime(0)\sim -690$ MeV (in case of the Skyrme term-stabilized
soliton solution), Eq.(\ref{PMTS}) implies $g$-value as
$g\approx -0.61$ which is comparable to that of the non-relativistic
quark model ($-0.75$) and the experimental estimation via the
$D^*$-decay
($|g|^2\raisebox{-0.6ex}{$\stackrel{\textstyle <}{\sim}$} 0.5)$.
This set of parameters yields a prediction on the
the $\Lambda_b$ mass and the average mass of the
$\Sigma_{c}^{}$-$\Sigma^*_{c}$ multiplets, $\overline{m}_{\Sigma_c^{}}
(\equiv\frac13(2 m_{\Sigma^*_c}+m_{\Sigma_c}^{}))$ as
\begin{equation}
\renewcommand{\arraystretch}{1.5} \begin{array}{l}
m_{\Lambda_b}^{} = M_{sol} + \overline{m}_B
                 - \frac32gF^\prime(0) + 3/8{\cal I}
                 = 5623 \mbox{ MeV}, \\
\overline{m}_{\Sigma_c^{}} = M_{sol} + \overline{m}_D
            - \frac32gF^\prime(0) + 11/8{\cal I}
            = 2483 \mbox{MeV},
\end{array}
\end{equation}
which are comparable with the experimental value of the $\Lambda_b$
mass (5641 MeV) and $\Sigma_c$ mass (2453 MeV)\cite{PDG}.
Recent experimental data for the $\Sigma_c^*$ mass of
2530~MeV\cite{SKAT} (although it needs the verifications of
other groups) gives 2504 MeV for the experimental value of
${\overline m}_{\Sigma_c}$, which is not far from our estimation.
In our approach, the $\Sigma_c$ and $\Sigma_c^*$ are degenerate in
mass. To get the splitting between them, one needs to include
$1/m_Q$ corrections\cite{OPM}.

With $1/{\cal I} \sim 200$ MeV, we can estimate the discrepancy
in the masses of the heavy baryons given by the mass formulas,
(\ref{MF}) and (\ref{MF1}), as $3/8{\cal I}\sim 60$ MeV. It
amounts 10\% of the binding energy $-\frac32gF^\prime(0)\sim 630$ MeV
and 30\% of the rotational energy $\sim 1/{\cal I} \sim 200$ MeV.
One should {\em not} say that the discrepancy is at most 3\% for the
case of the charmed baryons by comparing it with the whole
heavy baryon masses. Although we are working with the heavy baryons,
our scheme is valid only in the low energy region below
$\Lambda_{\mbox{\small QCD}}$.

So far, we have considered the first order perturbations in the
masses of the bound $|i,j_\ell^\pi\rangle\!\rangle$ states.
To estimate naively the effects of the other states, we take into
account the positive energy states into our procedure.
It leads to a mass correction
to the heavy baryon described by $|i,j^\pi_\ell\rangle\!\rangle_a$ as
\begin{equation}
\Delta m_{(i,j_\ell^\pi)}^{} = \sum_{b}
\frac{|{^{}_a\langle\!\langle
i,j_\ell^\pi|{\tilde H}_{rot}|i,j_\ell^\pi\rangle\!\rangle_b^{}}|^2}
{\varepsilon_a-\varepsilon_b}+\cdots,
\label{MassCor}
\end{equation}
where the summation runs over the unbound states
$|i,j_\ell^\pi\rangle\!\rangle_b$. They are, thus, at most of
second order in $1/N_c$, which is out of our concern. Here, to
check a consistency, we evaluate the leading order correction
to $(i=1,j_\ell^\pi=0^-)$ heavy baryon state. We have
\begin{equation}
\Delta m_{(1,0^-)}^{} = -
\frac{|{^{}_1\langle\!\langle 1,0^-| {\tilde H}_{rot}
|1,0^-\rangle\!\rangle_2^{}}|^2}{2gF^\prime(0)} =
-\frac{1}{2gF^\prime(0)}\frac{1}{18{\cal I}^2}
\sim -2.6 \mbox{ MeV},
\end{equation}
which is negligibly small compared to the first order corrections,
$11/8{\cal I}\sim 270$ MeV. The coupling of $1/N_c$ order due to
collective rotations cannot compete with the energy difference
$2gF^\prime(0)$ of order $N_c^0$.


\section{Pentaquark Exotic Baryons}

In the limit of infinite heavy quark mass, the quark model predicts
stable pentaquark($P$) exotic baryons whose quark contents are
$\bar{Q}q^4$\cite{Lipkin,GSR}. With the exact $SU(3)_F$ symmetry
assumed for the light quarks, it was shown that a strange
anti-charmed baryon $P_{\bar{c}s}^{}$
($\bar{c}sq_0^3$, $q_0=u,d$) is stable against the decays
into $\Lambda D$ or $ND_s$ with binding energy about 150 MeV.
The binding energy becomes down to $\sim 85$ MeV if included a
realistic $SU(3)_F$ symmetry breaking\cite{KZ} and it becomes even
unbound when the motions of the heavy constituent are taken into
account\cite{FGRS,ZR,TNK}. In the Skyrme model, as the heavy meson
masses increases, there appears bound state(s) for the antiflavored
heavy mesons to the soliton, which reveals a possibility for the stable
nonstrange $P$-baryon(s)\cite{RS93}. It is interesting to note that
in quark model such a nonstrange anticharmed baryon ($\bar{c}q_0^4$)
cannot have sufficient symmetry to yield a hyperfine binding.

Our approach can be easily switched to the one for the
soliton-antiflavored heavy meson bound system by considering
the negative energy solutions with the four velocity $v_\mu$
in the equations replaced by $-v_\mu$. (One may develop an
effective Lagrangian proper for the antiflavored heavy mesons.
See Ref.\cite{OPM0}.) Now, for each $k_\ell(\neq 0)$, we have
{\em three} degenerate bound states of the antiflavored heavy
mesons, $|k_\ell\rangle_{1,2,+}$, with the binding energy
$\frac12gF^\prime(0)$. Compared with that for the heavy mesons,
the binding energy is reduced by a factor 3. With
$gF^\prime(0)=419$ MeV as given by Eq.(\ref{PMTS}), it amounts
to $\sim 210$ MeV and is comparable to $1/{\cal I}$.

In Table III listed are the {\em bound}
$|i,j_\ell^\pi\rangle\!\rangle$ states proper for the discussions
of the $P$-baryons. As for the states with quantum numbers
$(i=0,j_\ell^\pi=0^-)$, $(1,0^-)$ and $(0,1^-)$, we have only
one bound state. To first order in $1/N_c$, the masses of such
$P$-baryons are given by the same mass formula as Eq.(\ref{MF1}):
\begin{equation}
m_{(i,j_\ell^\pi)}^P = M_{sol} + \omega_B^\prime
    + \frac{1}{2{\cal I}} \{cj_\ell(j_\ell+1) + (1-c) i(i+1)
    - ck_\ell(k_\ell+1) +\textstyle\frac34\},
\label{MF3}
\end{equation}
with $\omega^\prime_B=\overline{m}_\Phi^{}-\frac12gF^\prime(0)$.
The $c$-factors associated with $|0\rangle_1$ and $|1\rangle_+$ states
are obtained as $0$ and $-1/4$, respectively.\cite{fn1} The negative
$c$-value is remarkable. However, these states are associated with
the zero rotor-spin state and such nonvanishing $c$-values do not
play any important role in their masses which are simply
obtained as
\begin{equation}
m_{(0,0^-)}^{} = m_{(0,1^-)}^{}
= M_{sol} + \omega^\prime_B + \frac{3}{8{\cal I}}.
\end{equation}
For the case of $(1,0^-)$, the mass formula (\ref{MF3}) gives
\begin{equation}
m_{(1,0^-)} = M_{sol} + \omega^\prime_B + \frac{15}{8{\cal I}}
 = M_N + {\overline m}_\Phi - \frac12 g F'(0) + \frac{3}{2{\cal I}}.
\end{equation}
Since $\textstyle\frac12 gF^\prime(0)\sim 210$ MeV and
$1/{\cal I} \sim 195$ MeV, the rotational energy blows up the state
above the decay threshold of the nucleon-heavy meson bound system,
$M_{th}(\equiv M_N^{}+\overline{m}_\Phi)$.

In cases of the $(0,1^+)$ and $(1,0^+)$, we have two degenerate
states $|0,1^+\rangle\!\rangle_{1,2}$ and
$|1,0^+\rangle\!\rangle_{1,2}$, respectively. Since the states
come from the single-particle Fock states of the same $k_\ell$,
the energy matrices can be expressed in a form of Eq.(\ref{MF3})
with the constant $c$ replaced by the $2\times2$ matrix associated
with $k_\ell=1$:
\begin{equation}
c = \textstyle\frac{1}{4}\displaystyle\left(\!\! \begin{array}{cc}
1\!\!&\!\! 0  \\ 0 \!\!&\!\! -2
\end{array}\!\!\right).
\label{c2x2}
\end{equation}
Explicitly, we have
\begin{equation}
{\cal E}_{(0,1^+)} = M_{sol} + \omega^\prime_B
+ \displaystyle\frac{3}{8{\cal I}},
\hskip 5mm \mbox{and} \hskip 5mm
{\cal E}_{(1,0^+)} = M_{sol} + \omega^\prime_B
+ \displaystyle\frac{11}{8{\cal I}}
-\frac{2c}{{\cal I}}.
\label{EM10}
\end{equation}
Here again, the energy matrix ${\cal E}_{(0,1^+)}$ is independent
of $c$ because the states are made of the zero rotor-spin state.
Anyway, both are diagonal so that $|0,1^+\rangle\!\rangle_{1,2}$
and $|1,0^+\rangle\!\rangle_{1,2}$ are the
eigenstates of the Hamiltonian with the masses
\begin{equation}
m_{(0,1^+)_{1,2}} = M_{sol} + \omega^\prime_B + \frac{3}{8{\cal I}} = M_N +
\overline{m}_\Phi - \displaystyle \frac12 gF^\prime(0),
\end{equation}
and
\begin{equation}
\renewcommand{\arraystretch}{1.7} \begin{array}{l}
m^{}_{(1,0^+)_1} = M_{sol} + \omega^\prime_B +
\displaystyle\frac{7}{8{\cal I}}
= M_N + \overline{m}_\Phi - \displaystyle\frac{1}{2}
gF^\prime(0) + \frac{1}{2{\cal I}}, \\
m^{}_{(1,0^+)_2} = M_{sol} + \omega^\prime_B +
\displaystyle\frac{19}{8{\cal I}}
= M_N + \overline{m}_\Phi - \displaystyle\frac{1}{2}
gF^\prime(0) + \frac{2}{{\cal I}}.
\end{array}
\end{equation}
The state $|1,0\rangle\!\rangle_{2}$ lies above
the decay threshold.

For the $(i=1, j^\pi_\ell=1^+)$ states, we have three possible
combinations; $|1,1^+\rangle\!\rangle_{1,2,3}$. As far as the
first two states $|1,1^+\rangle\!\rangle_{1,2}$ are concerned,
the energy matrix reads simply
\begin{equation}
{\cal E}_{(1,1^+)} = M_{sol} + \omega^\prime_B
+ \displaystyle\frac{11}{8{\cal I}} -\frac{c}{{\cal I}},
\end{equation}
with the same $2\times2$ matrix $c$ as in Eq.(\ref{c2x2}). It leads
us to the mass eigenvalues
\begin{equation}
\renewcommand{\arraystretch}{1.7} \begin{array}{l}
m^{}_{(1,1^+)_1} = M_{sol} + \omega^\prime_B +
\displaystyle\frac{9}{8{\cal I}}, \\
m^{}_{(1,1^+)_2} = M_{sol} + \omega^\prime_B +
\displaystyle\frac{15}{8{\cal I}}.
\end{array}
\end{equation}
With including the third state $|1,1^+\rangle\!\rangle_3$,
the full $3\times3$ energy matrix is obtained as
\begin{equation}
{\cal E}_{(1,1^+)} = M_{sol} + \omega^\prime_B +
\frac{11}{8{\cal I}} + \frac{1}{4{\cal I}}\left(\!\!
\renewcommand{\arraystretch}{1} \begin{array}{ccc}
  -1   \!&\! 0 \!&\! \sqrt3 \\
   0   \!&\! 2 \!&\!   0 \\
\sqrt3 \!&\! 0 \!&\!   1
\end{array} \!\!\right).
\end{equation}
The coupling between $|1,1^+\rangle\!\rangle_{1}$ and
$|1,1^+\rangle\!\rangle_3$ modifies $m^{}_{(1,1^+)_1}$ as
\begin{equation}
m^-_{(1,1^+)} = M_{sol} + \omega^\prime_B +
\displaystyle\frac{7}{8{\cal I}},
\end{equation}
with $|1,1^+\rangle\!\rangle_- = \frac{\sqrt3}{2} |1,1^+\rangle\!\rangle_1
- \frac12 |1,1^+\rangle\!\rangle_3 $
and adds doubly degenerate unbound states of mass
$M_{sol}+\omega^\prime_B + 15/8{\cal I}$.
Note that in obtaining those off-diagonal matrix elements
one cannot use Eq.(\ref{WE1}) and they cannot be written
in a form of Eq.(\ref{MF3}) anymore.

As for the $|1,1^-\rangle\!\rangle$ states, we have four
possible combinations. They are made of the heavy-meson bound
state with different $k_\ell$(=0,1,2). By using Eq.(\ref{WE}),
we obtain the energy matrix as
\begin{equation}
{\cal E}_{(1,1^-)} = M_{sol} + \omega^\prime_B +
\frac{11}{8{\cal I}} + \frac{1}{12{\cal I}} \left(\!\!
\renewcommand{\arraystretch}{1} \begin{array}{cccc}
   0    \!\!&\!\! 4\sqrt3   \!\!&\!\! 0 \!\!&\!\! 0 \\
4\sqrt3 \!\!&\!\!    3      \!\!&\!\! 0 \!\!&\!\! \sqrt{15} \\
   0    \!\!&\!\!    0      \!\!&\!\! -6 \!\!&\!\! 0 \\
   0    \!\!&\!\! \sqrt{15} \!\!&\!\! 0 \!\!&\!\! 9
\end{array} \!\!\right).
\end{equation}
The couplings between the states are somewhat strong.
It yields four mass eigenenergies
\begin{equation}
\renewcommand{\arraystretch}{1.7} \begin{array}{l}
m^-_{(1,1^-)} = M_{sol} + \omega^\prime_B +
\displaystyle\frac{7}{8{\cal I}},\\
m^{1,+}_{(1,1^-)} = M_{sol} + \omega^\prime_B +
\displaystyle\frac{15}{8{\cal I}}, \\
m^{2,+}_{(1,1^-)} = M_{sol} + \omega^\prime_B +
\displaystyle\frac{19}{8{\cal I}},
\end{array}
\end{equation}
where the states with $m_{(1,1^-)}^-$ are doubly degenerate
and expected to be bound.

Our remarkably simple results on the $P$-baryon masses are
summarized in Table IV and Fig.~3. To provide a rough scale,
in Fig.~3, the heavy baryon spectrum is presented in the left hand
side. Here again, the degeneracy of the $P$-baryon states in the
parity is apparent. In Table IV, we also give a rough prediction
obtained by using the values for the parameters given by
Eq.(\ref{PMTS}). However, all the states listed in Table IV do not
seem to survive under the finite heavy meson mass corrections.
Recently, we have reported that such finite mass corrections reduce
the binding energy by an amount from 25\% (in case of bottomed
baryons) to 35\% (in case of the charmed baryons) of their
infinite mass limit, $\frac32gF^\prime(0)$\cite{OPM}. Note that
35\% of $\frac32gF^\prime(0)$ is comparable to the binding energy
$\frac12gF^\prime(0)$ for the soliton-antiflavored heavy mesons.
The finite mass corrections will be more crucial for the heavy meson
eigenstates with $\ell_{\mbox{\scriptsize eff}}\neq 0$.
All the degeneracies in the heavy meson eigenstates will be broken.
Then, the couplings between the states as represented by the energy
matrices ${\cal E}_{(1,0^+)}$, ${\cal E}_{(1,1^-)}$ and
${\cal E}_{(1,1^+)}$ will do less important roles. However, it does
not mean that one can apply the mass formula Eq.(\ref{MF}) for the
$P$-baryons in the present form.

In order to show it, here, we compare our results with what would
have been obtained by straightforwardly extending the bound state
approach of CK as in Ref.\cite{RS93}. As a nontrivial example, we
consider the $P$-baryons with $i=1$ and
$j^\pi=\frac12^+,\frac32^+(j^\pi_\ell=0^+,1^+)$, which are obtained
by combining the rotor-spin state with $i=1$ to the heavy meson bound
states with $k_\ell^\pi=1^+$. To simplify the process, we will include
only the $|1\rangle_1$ state into the consideration, in which case
the $P$-baryon masses are simply given by the diagonal
elements of the energy matrices ${\cal E}_{(1,0^+)}$ or
${\cal E}_{(1,1^+)}$ associated with $|1,0^+\rangle\!\rangle_1$ or
$|1,1^+\rangle\!\rangle_1$ :
\begin{equation}
\renewcommand{\arraystretch}{1.5} \begin{array}{l}
m_{(1,\frac12^+)} = M_{sol} + \omega^\prime_B
+ (11-16 c_1)/8{\cal I}, \\
m_{(1,\frac12^+)} = m_{(1,\frac32^+)} =
M_{sol} + \omega^\prime_B + (11-8 c_1)/8{\cal I},
\label{Mtest}
\end{array}
\end{equation}
with $ c_1=\frac14$, the $c$-value associated with the state
$|1\rangle_1$.  In the traditional bound state approach with
sufficiently heavy meson masses, $|1\rangle_2$ will appear as nearly
doubly degenerate states, say, $|k=\frac12\rangle$ and
$|k=\frac32\rangle$, which become to completely degenerate
in the infinitely heavy mass limit. When combined with $i$=1 rotor
spin state, the former yields $P$-baryons of
$j^\pi=\frac12^+, \frac32^+$ with masses
\begin{equation}
\renewcommand{\arraystretch}{1.6} \begin{array}{l}
m^{}_{(1,\frac12^+)} = M_{sol} + \omega^\prime_B +
(11 - 8c_{\frac12}^{})/8{\cal I}, \\
m^{}_{(1,\frac32^+)} = M_{sol} + \omega^\prime_B +
(11 + 4c_{\frac12}^{})/8{\cal I},
\end{array}
\end{equation}
and the latter yields
\begin{equation}
\renewcommand{\arraystretch}{1.6} \begin{array}{l}
m^{}_{(1,\frac12^+)} = M_{sol} + \omega^\prime_B +
(11 - 20c_{\frac32}^{})/8{\cal I}, \\
m^{}_{(1,\frac32^+)} = M_{sol} + \omega^\prime_B +
(11 - 8c_{\frac32}^{})/8{\cal I}.
\end{array} \label{MFours}
\end{equation}
Here, $c_{\frac12}$ and $c_{\frac32}$
are the $c$-values associated with the $|k=\frac12\rangle$ and
$|k=\frac32\rangle$ states. These $c$-values are obtained as
\begin{equation}
\textstyle c_{\frac12} = \frac43 c_1 \hskip 5mm
\mbox{and} \hskip 5mm c_{\frac32} = \frac23 c_1.
\label{MFCK}
\end{equation}
Note that the $P$-baryon masses given by Eqs.(\ref{MFCK})
are quantitatively different from those of Table~IV and
furthermore that they violate the heavy quark symmetry; that is,
there is no signal for any degenerate pairs of $(1,\frac12^+)$
and $(1,\frac32^+)$.

Then, what goes wrong in this straightforward extension? In the
quantization procedure for obtaining Eq.(\ref{MF}), only a single
bound state is involved. Thus, to obtain the hyperfine energy of order
$1/N_c$, it is enough to take the expectation value of
${\tilde H}_{rot}$ with respect to the unperturbed soliton-heavy meson
bound state of order $N_c^0$. However, in the heavy meson mass limit,
all the heavy meson bound states come in degenerate doublets with
grand spin $k=k_\ell\pm 1/2$ (unless $k_\ell=0$) and consequently
we have degenerate states up to order $m_Q^0 N_c^0$, for example,
$|i=1,j=\frac12\rangle\!\rangle$ coming from $k=\frac12$ bound state
and $|i=1,j=\frac12\rangle\!\rangle$ from $k=\frac32$ bound state.
In evaluating the hyperfine energy of next order due to collective
rotation, as is well-known in standard quantum mechanics, we have
to diagonalize the energy matrices for the degenerate basis. As for
our sample case, we obtain the energy matrices as
\begin{equation}
\renewcommand{\arraystretch}{1.5} \begin{array}{l}
\displaystyle {\cal E}_{(1,\frac12^+)} =
M_{sol} +\omega^\prime_B + \frac{11}{8{\cal I}}
+ \frac{c_1}{6{\cal I}}\left(\!\! \renewcommand{\arraystretch}{1}
\begin{array}{cc} -8 \!&\! 2\sqrt2 \\  2\sqrt2 \!&\! -10
\end{array} \!\!\right), \\
\displaystyle {\cal E}_{(1,\frac32^+)} =
M_{sol} + \omega^\prime_B + \frac{11}{8{\cal I}}
+ \frac{c_1}{6{\cal I}}\left(\!\!
\renewcommand{\arraystretch}{1} \begin{array}{cc}
4 \!&\! -2\sqrt5 \\  -2\sqrt5 \!&\! -4
\end{array} \!\!\right),
\end{array}
\end{equation}
which yield the $P$-baryon masses consistent with Eq.(\ref{Mtest}) as
\begin{equation}
\renewcommand{\arraystretch}{1.6} \begin{array}{l}
m_{(1,\frac12^+)} = M_{sol} + \omega^\prime_B
+ (11 - 16c_1)/8{\cal I}, \\
m_{(1,\frac12^+)} = M_{sol} + \omega^\prime_B
+ (11 - 8c_1)/8{\cal I}, \\
m_{(1,\frac32^+)} = M_{sol} + \omega^\prime_B
+ (11 - 8c_1)/8{\cal I}, \\
m_{(1,\frac32^+)} = M_{sol} + \omega^\prime_B
+ (11 + 8c_1)/8{\cal I}.
\end{array}
\end{equation}

It shows that the inclusion of the degenerate states into the
quantization procedure is, thus, essential in restoring the heavy
quark symmetry in the bound state approach. If we work with finite
but sufficiently heavy meson masses, we have only approximate
degenerate states of the grand spin $k = k_\ell \pm \frac12$.
Although the mass corrections due to the nondegenerate states are at
most of order $1/N_c^2$, because of the small energy discrepancy of
order $1/m_Q$ in the denominator, their couplings can compete with
diagonal terms of order $m_Q^0 N_c^{-1}$ and should be
taken into account properly.


\section{Summary and Conclusion}

In this paper, we have discussed the heavy quark symmetry in describing
heavy baryons containing a single heavy quark or antiquark as bound
states of the $SU(2)$ soliton and heavy mesons. We have developed a
consistent bound state approach so that the heavy quark symmetry is
realized explicitly in the heavy baryon spectrum in the infinitely
heavy mass limit. The resulting mass formula reads
\begin{equation}
\renewcommand{\arraystretch}{1.7} \begin{array}{ll}
m_{(0,j^\pi)} = M_{sol} + \omega_B +
\displaystyle \frac{3}{8{\cal I}}, &
( j^\pi = \frac12^\pm, \frac32^- ) \\
m_{(1,j^\pi)} = M_{sol} + \omega_B +
\displaystyle \frac{11}{8{\cal I}}, &
( j^\pi = \frac12^\pm, \frac32^\pm) \\
\end{array}
\end{equation}
for the heavy baryons and
\begin{equation}
\renewcommand{\arraystretch}{1.7} \begin{array}{ll}
m_{(0,j^\pi)} = M_{sol} + \omega_B^\prime
+ \displaystyle \frac{3}{8{\cal I}}, &
( j^\pi = \frac12^\pm, \frac32^\pm) \\
m_{(1,j^\pi)} = M_{sol} + \omega_B^\prime
+ \displaystyle \frac{7}{8{\cal I}}, &
( j^\pi = \frac12^\pm, \frac32^\pm) \\
\end{array}
\end{equation}
for the $P$-baryons. All the masses are consistent with the heavy
quark symmetry and the genuine hyperfine splittings vanish.
As for the heavy baryons, one can still apply the mass formula
(\ref{MF}) to obtain their masses and the $c$-factor defined by
Eq.(\ref{cfactor}) plays the role of the hyperfine constant.
For the $P$-baryons, the degenerate states up to order
$m_Q^0 N_c^0$ requires diagonalizing the Hamiltonian with respect
to the degenerate basis to obtain the hyperfine energy of next
order in $1/N_c$. In this case, the nonvanishing $c$-factors
defined by Eq.(\ref{cfactor}) should not be treated as the
hyperfine constant. In restoring the heavy quark symmetry in
the $P$-baryon spectrum, the couplings of the states made of the
degenerate heavy meson bound states with the grand spins
$k=k_\ell\pm 1/2$ are shown to play an important role.

Except the ground state with $i$=0 and $j^\pi$=$\frac12^+$, the
occurrence of the parity doublets in the spectrum ({\it i.e.\,},
two states with the same angular momentum but opposite parity
occurring at the same mass) is interesting. However, we are not in
the position to conclude whether such a parity doubling has any
physical importance as discussed in the context of chiral
symmetry\cite{ChS} and Regge-pole theory\cite{RPT} or just an
artifact from our approximation on the radial functions.
To extract more decisive conclusion, we should work with finite
heavy mesons incorporating the kinetic terms\cite{OPM2}
and see whether the parity doubling occurs in the heavy quark limit.
Iachello\cite{Ia} has reported that a similar parity doubling in the
excited baryon spectra is occurred in the baglike models
and stringlike models and it was analyzed as a consequence
of the geometric structure of baryons.

We have corrected a mistake committed in applying the traditional
bound state approach of evaluating the expectation value of the
operator $\vec\Theta^2$ to the heavy baryons. It has been
approximated simply by the square of the expectation value of
$\vec\Theta$ as
\begin{equation}
\langle\!\langle i,j_\ell | \vec\Theta^2 | i,j_\ell\rangle\!\rangle
\approx c^2_{} k_\ell ( k_\ell + 1 ).
\end{equation}
In our approach,
it can be exactly obtained as
\begin{equation}
\langle\!\langle i,j_\ell | \vec\Theta^2 |
i,j_\ell\rangle\!\rangle = \textstyle\frac34.
\end{equation}
In case of heavy baryons with $c=0$, the correction
is an overall shift of the heavy baryon masses by an amount
$3/8{\cal I} \sim 60$ MeV.

Furthermore, the simple structure of the model enables us to
illustrate explicitly how the spin-grandspin transmutation occurs
in the bound state approach. In the isospin co-moving frame,
the total spin of the soliton-heavy meson bound system
can be obtained as
\begin{equation}
{\vec J}={\vec R} + \vec{K}_{{\mbox{\scriptsize bf}}},
\end{equation}
with the rotor spin ${\vec R}$ and the grand spin
$\vec{K}_{{\mbox{\scriptsize bf}}}$
of the heavy meson fields in the isospin co-moving frame.
The latter plays the role of the heavy-meson spin; that is,
the isospin of the heavy mesons is transmuted to the part of their
spin.

We have worked with infinitely heavy mesons (and soliton).
It provides a useful instruction to the bound state approaches with
finite-mass heavy mesons so that it has a correct heavy quark limit;
that is, one has to include the nearly doubly degenerate heavy meson
bound states of grand spin $k=k_\ell\pm 1/2$ into the quantization
procedure.

\acknowledgements
We are grateful to Mannque Rho and Hyun Kyu Lee for fruitful
discussions. The work of YO was supported by the National Science
Council of ROC under grant No. NSC83-0208-M002-017. BYP and DPM were
supported in part by the Korea Science and Engineering
Foundation through the Center for Theoretical Physics, SNU.

\appendix


\section{Spin and Isospin Operators and Their Eigenstates}

The invariance of the Lagrangian density (\ref{Lfree}) under the
infinitesimal Lorentz transformation
\begin{equation}
\renewcommand{\arraystretch}{1.3} \begin{array}{l}
x^\mu \rightarrow x^{\prime\mu} = x^\mu + \epsilon^\mu_\nu x^\nu,
\mbox{ with $\epsilon^{\mu\nu}=-\epsilon^{\nu\mu}$ } \\
\Phi \rightarrow \Phi^{\prime}(x^\prime) = \Phi(x), \\
\Phi^*_\alpha(x) \rightarrow \Phi^{\prime*}_\alpha(x^\prime)
= \frac12\epsilon^{\mu\nu}{(S_{\mu\nu})_\alpha}^\beta
\Phi^*_\beta(x), \mbox{ with $(S^{\mu\nu})_{\alpha\beta} =
{g^\mu}_\alpha {g^\nu}_\beta - {g^\mu}_\beta {g^\nu}_\alpha$ }
\end{array}
\end{equation}
defines conserved angular momentum operators
\begin{equation}
\renewcommand{\arraystretch}{1.5} \begin{array}{l}
J^i = \textstyle\frac12 \varepsilon^{ijk}
\displaystyle\int\! d^3r {\cal M}_{0jk}, \\
{\cal M}^{0jk} = \left( x^j {\cal P}^{0k} -
x^k {\cal P}^{0j} \right)
+ \left( \Pi^{*m} (S^{kj})_{mn}  \Phi^{*n\dagger}
+ \Phi^{*m} (S^{kj})_{mn} \Pi^{*n\dagger} \right)
\end{array}
\end{equation}
where ${\cal P}^{\mu\nu}$ is the canonical energy-momentum tensor
and $\Pi^{*n} (\equiv \partial{\cal L}_{\mbox{\scriptsize free}}/
\partial\dot{\Phi}^*_n)$ is the
momentum conjugate to the field $\Phi^{*n}$.
Here, the indices run from 1 to 3. The first part corresponds to
the orbital angular momentum, ${\vec L}$, and the second part
correspond to the spin angular momentum, ${\vec S}$. In the heavy
meson mass limit ($m_{\Phi,\Phi^*}^{}\rightarrow\infty$),
substitution of Eq.(\ref{H}a) with $v=(1,\vec{0})$
leads to the spin operator of the heavy meson fields as
\begin{equation}
{\vec S} = i \int\! d^3r \, \vec{\Phi}^*_v \times
\vec{\Phi}^{*\dagger}_v
\end{equation}
as the leading order term in the meson masses. From now on, we
will work in the rest frame of the heavy mesons. In terms of the
$4\times4$ matrix field $H(x)$, it can be simply rewritten as
\begin{equation}
{\vec S} = - \int\!\! d^3r \,\mbox{Tr}([\textstyle\frac12
{\vec\sigma}, H] \bar{H}),
\label{S}
\end{equation}
which implies that the corresponding quantum mechanical spin
operators in the $4\times4$ matrix representation is the Dirac
spin matrices acting on the wavefunction $H_n(x)$ as
\begin{equation}
{\vec S}\{H_n\} \equiv [\textstyle\frac12 {\vec\sigma}, H_n].
\end{equation}
The minus sign in Eq.(\ref{S}) is due to the normalization convention
of the field $H(x)$ and $\bar{H}(x)$. Note that the meson number
operator is given by
\begin{equation}
N = - \int\!\!d^3r\,\mbox{Tr}(H\bar{H}).
\end{equation}
One can easily find the eigenstates of the spin operators as
\begin{equation}
\renewcommand{\arraystretch}{1.3} \begin{array}{l}
(\mbox{$s$=0;$s_3$=0})=[\frac{1}{2\sqrt2}(1+\gamma^0)]\gamma_5, \\
(\mbox{$s$=1;$s_3$=0})=[\frac{1}{2\sqrt2}(1+\gamma^0)]\gamma^3, \\
(\mbox{$s$=1;$s_3$=$\pm$1})=[\frac{1}{2\sqrt2}(1+\gamma^0)]
  [ \frac{\mp1}{\sqrt2} (\gamma^1\pm i\gamma^2) ],
\end{array}
\end{equation}
which are normalized as
\begin{equation}
\,\mbox{Tr}\{ (s;s_3)\overline{(s^\prime;s^\prime_3)}\}
= -\delta_{s^\prime s} \delta_{s_3^{} s^\prime_3}.
\end{equation}

Furthermore, the invariance of the Lagrangian (\ref{Lfree2})
under the heavy-quark spin rotation (\ref{H}b) leads to
conserved heavy-quark spin operators ${\vec S}_Q$ as
\begin{equation}
{\vec S}_Q = - \int\!\! d^3r \,\mbox{Tr}
(\textstyle\frac12 {\vec\sigma} H \bar{H}),
\end{equation}
that is, the action of the heavy-quark spin operators on the
wavefunction is the multiplication of the Dirac spin matrices
to its left-hand-side:
\begin{equation}
{\vec S}_Q\{H_n\} = \textstyle\frac12 {\vec\sigma} H_n.
\end{equation}
Since the heavy-quark spin decouples in the heavy meson mass limit,
it is convenient to introduce the ``light-quark" spin operators for
the light degrees of freedom in the heavy mesons:
\begin{equation}
\renewcommand{\arraystretch}{1.5} \begin{array}{c}
{\vec S}_\ell (\equiv {\vec S} - {\vec S}_Q )
= + \displaystyle\int\!\! d^3r \,\mbox{Tr}
(H\textstyle\frac12 {\vec\sigma}\bar{H}), \\
{\vec S}_\ell\{H_n\} = H_n(-\textstyle\frac12 {\vec\sigma}).
\end{array}
\end{equation}
The eigenstates of ${\vec S}_Q$ and ${\vec S}_\ell$ can be explicitly
obtained as
\begin{equation}
\renewcommand{\arraystretch}{1.3} \begin{array}{l}
^{}_{\ell}( {+\textstyle\frac12 |}{|+\textstyle\frac12 )_Q^{}}
= (\mbox{$s$=1;$s_3$=$+$1}), \\
^{}_{\ell}( {+\textstyle\frac12 |}{|-\textstyle\frac12 )_Q^{}}
= \frac{1}{\sqrt2}\{
  (\mbox{$s$=1;$s_3$=0}) + (\mbox{$s$=0;$s_3$=0}) \}, \\
^{}_{\ell}( {-\textstyle\frac12 |}{|+\textstyle\frac12 )_Q^{}}
= \frac{1}{\sqrt2}\{
  (\mbox{$s$=1;$s_3$=0}) - (\mbox{$s$=0;$s_3$=0}) \}, \\
^{}_{\ell}( {-\textstyle\frac12 |}{|-\textstyle\frac12 )_Q^{}}
= (\mbox{$s$=1;$s_3$=$-$1}).
\end{array}
\end{equation}

On the other hand, the isospin operators associated with the
invariance of the Lagrangian (\ref{LHQS}) under the isovector
transformations (\ref{ct1}) and (\ref{ct2}) (with $L$=$R$=$U$) are
\begin{equation}
{\vec I} = {\vec I}_M + {\vec I}_h,
\label{I}
\end{equation}
with ${\vec I}_M$ the isospin operator of the Goldstone boson fields
$$
{\vec I}_M = i \int\!\! d^3r \frac{f^2_\pi}{2}\,
\mbox{Tr}\{\textstyle\frac12 {\vec\tau}
(\Sigma^\dagger\partial_0 \Sigma + \Sigma\partial_0 \Sigma^\dagger)\}
+\cdots,
\eqno(\mbox{\ref{I}a})
$$
and ${\vec I}_h$ those of the heavy meson fields interacting with Goldstone
pions
$$
{\vec I}_h = + \textstyle \frac12 \displaystyle \int\!\! d^3r \,\mbox{Tr}
\{ H \textstyle\frac12 (\xi^\dagger \vec\tau \xi + \xi \vec\tau \xi^\dagger
)\bar{H} \}.
\eqno(\mbox{\ref{I}b})
$$
Especially, the isospin operator of the free heavy meson fields is
\begin{equation}
{\vec I}_h = - \int\!\! d^3r \,\mbox{Tr}
\{ H(-\textstyle\frac12 {\vec\tau})\bar{H} \}.
\end{equation}
Thus, the quantum mechanical isospin operator for the wavefunction
is the $2\times2$ Pauli matrices $(-\textstyle\frac12 {\vec\tau})$
acting on the right-hand-side of the anti-isodoublet $H_n(x)$:
\begin{equation}
{\vec I}_h\{H_n(x)\} = H_n(x)(-\textstyle\frac12 {\vec\tau}).
\end{equation}
Explicitly, their eigenstates can be written as
\begin{equation}
\tilde{\phi}_{+\frac12} = (0, -1), \hskip 5mm \mbox{and}
\hskip 5mm  \tilde{\phi}_{-\frac12} = (1,0).
\end{equation}


\section{$K_\ell$-basis}

Here, we present the explicit expressions of the the angular part
of the wavefunctions, which are the eigenstates of $K_\ell^2$,
$K_{\ell,z}$ and $S_Q^2$, $S_{Q,z}$. They can be obtained by linear
combinations of direct products of four angular momentum bases;
$Y_{\ell m}({\hat r})$(the eigenstate of $L^2$, $L_z$),
$\tilde{\phi}_{\pm\frac12}$(the eigenstate of $I_h^2$, $I_{h,z}$),
$^{}_{\ell}( {\pm\frac12|}$(the eigenstate of $S_\ell^2$,
$S_{\ell,z}$), and ${|\pm\frac12)_Q^{}}$(the eigenstate of $S_Q^2$,
$S_{Q,z}$). For a given quantum numbers, $k_\ell,k_3$, we have four
states $K^{(i)}_{k_\ell k_3 s_Q}$, $i$=1,2,3,4 according to the
numbers of the different combination of $I_h$ and $S_\ell$.

We first combine the spherical harmonics $Y_{\ell m}({\hat r})$,
and the isospin basis $\tilde{\phi}_{\pm\frac12}$ to obtain
${\cal Y}_{\lambda\lambda_3}^{\pm}({\hat r})$, the eigenstates
of $\Lambda^2$ and $\Lambda_3$ ($\vec{\Lambda} \equiv {\vec L} +
{\vec I}_h$):
\begin{equation}
\renewcommand{\arraystretch}{1.6} \begin{array}{l}
\displaystyle{\cal Y}^{(+)}_{\lambda\lambda_3}
= \sqrt{\frac{\lambda + \lambda_3}{2\lambda}}
  Y_{\ell\, \lambda_3-\frac12}({\hat r}) \tilde{\phi}_{+\frac12}
+ \sqrt{\frac{\lambda - \lambda_3}{2\lambda}}
  Y_{\ell\, \lambda_3+\frac12}({\hat r}) \tilde{\phi}_{-\frac12},
\mbox{\hskip 5mm $\lambda = \ell + \frac12$} \\
\displaystyle{\cal Y}^{(-)}_{\lambda\lambda_3}
= -\sqrt{\frac{\lambda - \lambda_3 + 1}{2(\lambda+1)}}
  Y_{\ell\, \lambda_3-\frac12}({\hat r}) \tilde{\phi}_{+\frac12}
+ \sqrt{\frac{\lambda + \lambda_3 + 1}{2(\lambda+1)}}
  Y_{\ell\, \lambda_3+\frac12}({\hat r}) \tilde{\phi}_{-\frac12},
\mbox{\hskip 5mm $\lambda = \ell - \frac12$}
\end{array}
\end{equation}
It provides a convenient basis in evaluating the expectation values of
the operator including $({\vec\tau}\!\cdot\!{\hat r})$, since
${\cal Y}^{(\pm)}$ satisfy a useful identity
\begin{equation}
{\cal Y}^{(\pm)}_{\lambda\,\lambda_3}({\vec\tau}\!\cdot\!{\hat r})
= {\cal Y}^{(\mp)}_{\lambda\,\lambda_3}.
\end{equation}

Next, the eigenstates ${\cal K}^{(i)}_{k_\ell k_3 s_Q}$ is obtained by
combining ${\cal Y}^{(\pm)}$ and $^{}_{\ell}(
{\pm\textstyle\frac12 |}{|\pm\textstyle\frac12 )_Q^{}}$:
\begin{equation}
\renewcommand{\arraystretch}{1.9} \begin{array}{l}
\mbox{ \hskip -1cm (i) $i$=1; $\lambda$=$\ell+\textstyle\frac12 $,
$k_\ell$=$\lambda-\frac12$=$\ell$ } \\
{\cal K}^{(1)}_{k_\ell k_3 s_Q}
= - \displaystyle\sqrt{\frac{k_\ell-k_3+1}{2(k_\ell+1)}}
    {\cal Y}^{(+)}_{\lambda k_3-\frac12}
    {^{}_{\ell}( {+\textstyle\frac12 |}{|s_Q)_Q^{}}}
  + \displaystyle\sqrt{\frac{k_\ell+k_3+1}{2(k_\ell+1)}}
    {\cal Y}^{(+)}_{\lambda k_3+\frac12}
    {^{}_{\ell}( {-\textstyle\frac12 |}{|s_Q)_Q^{}}}, \\
\mbox{ \hskip -1cm (ii) $i$=2; $\lambda$=$\ell-\textstyle\frac12 $,
$k_\ell$=$\lambda+\frac12$=$\ell$ } \\
{\cal K}^{(2)}_{k_\ell k_3 s_Q}
= + \displaystyle\sqrt{\frac{k_\ell+k_3}{2k_\ell}}
    {\cal Y}^{(-)}_{\lambda k_3-\frac12}
    {^{}_{\ell}( {+\textstyle\frac12 |}{|s_Q)_Q^{}}}
  + \displaystyle\sqrt{\frac{k_\ell-k_3}{2k_\ell}}
    {\cal Y}^{(-)}_{\lambda k_3+\frac12}
    {^{}_{\ell}( {-\textstyle\frac12 |}{|s_Q)_Q^{}}}, \\
\mbox{ \hskip -1cm (iii) $i$=3; $\lambda$=$\ell-\textstyle\frac12 $,
$k_\ell$=$\lambda-\frac12$=$\ell-1$ } \\
{\cal K}^{(3)}_{k_\ell k_3 s_Q}
= - \displaystyle\sqrt{\frac{k_\ell-k_3+1}{2(k_\ell+1)}}
    {\cal Y}^{(-)}_{\lambda k_3-\frac12}
    {^{}_{\ell}( {+\textstyle\frac12 |}{|s_Q)_Q^{}}}
  + \displaystyle\sqrt{\frac{k_\ell+k_3+1}{2(k_\ell+1)}}
    {\cal Y}^{(-)}_{\lambda k_3+\frac12}
    {^{}_{\ell}( {-\textstyle\frac12 |}{|s_Q)_Q^{}}}, \\
\mbox{ \hskip -1cm (iv) $i$=4; $\lambda$=$\ell+\textstyle\frac12 $,
$k_\ell$=$\lambda+\frac12$=$\ell+1$ } \\
{\cal K}^{(4)}_{k_\ell k_3 s_Q}
= + \displaystyle\sqrt{\frac{k_\ell+k_3}{2k_\ell}}
    {\cal Y}^{(+)}_{\lambda k_3-\frac12}
    {^{}_{\ell}( {+\textstyle\frac12 |}{|s_Q)_Q^{}}}
  + \displaystyle\sqrt{\frac{k_\ell-k_3}{2k_\ell}}
    {\cal Y}^{(+)}_{\lambda k_3+\frac12}
    {^{}_{\ell}( {-\textstyle\frac12 |}{|s_Q)_Q^{}}},
\end{array}
\end{equation}
They are normalized as
\begin{equation}
\int\! d\Omega \,\mbox{Tr}({\cal K}^{(i)}_{k_\ell k_3 s_Q}
\bar{{\cal K}}^{(i^\prime)}_{k_\ell^\prime k_3^\prime s^\prime_Q})
= - \delta_{ii^\prime} \delta_{k_\ell^{} k_\ell^\prime}
  \delta_{k^{}_3 k_3^\prime} \delta_{s_Q^{} s^\prime_Q}.
\label{Knorm}
\end{equation}
One can easily check that ${\cal K}^{(1)}_{k_\ell k_3 s_Q}$ and
${\cal K}^{(2)}_{k_\ell k_3 s_Q}$ have the parity $\pi=-(-1)^{k_\ell}$
and the other two have the parity $\pi=(-1)^{k_\ell}$ and they
are related to each other as
\begin{equation}
{\cal K}^{(i)}_{k_\ell k_3 s_Q} ({\vec\tau}\!\cdot\!{\hat r})
   = {\cal K}^{(i+2)}_{k_\ell k_3 s_Q}.
\mbox{\hskip 1cm ($i$=1,2)}
\label{Krel}
\end{equation}

These $K_\ell$-bases evaluate the matrix elements ${\cal M}_{ij}$
defined by Eq.(\ref{M}) as
\begin{equation}
\renewcommand{\arraystretch}{1.5} \begin{array}{c}
{\cal M}=-\textstyle\frac12gF^\prime(0)
\displaystyle \left(\!\!\begin{array}{cc}
 1 & 0  \\  0 & 1  \end{array}\!\!\right)
\mbox{\hskip 5mm (for $i,j$=1,2) } \hskip 2cm \\
\displaystyle {\cal M}=
\frac{\frac12gF^\prime(0)}{2k_\ell+1} \left(\!\!\begin{array}{cc}
2k_\ell+3 \!\!\!&\!\!\! -4\sqrt{k_\ell(k_\ell+1)} \\
-4\sqrt{k_\ell(k_\ell+1)} \!\!\!&\!\!\! 2k_\ell-1
\end{array}\!\!\right)
\mbox{ \hskip 5mm (for $i,j$=3,4) },
\end{array}
\end{equation}
independent of $k_3$ and the heavy quark spin $s^{}_Q$.

In Sec.~V, we work with the expectation values of the operator
${\vec\Theta}(\infty)$ with respect to the single-particle Fock
states $|n\rangle$ for which we need to evaluate the expectation
values of $\vec\Theta$ with respect to the $K_\ell$-basis.
Wigner-Eckart theorem enables us to express them as\cite{Edmonds}
\begin{equation}
{^{}_j\langle k'_\ell k'_3 | {\vec\Theta}^q(\infty)
|k_\ell k_3 \rangle_i^{}}
\equiv -\int\!\! d\Omega \,\mbox{Tr}
\{ {\cal K}^{(i)}_{k_\ell k_3 s_Q} \vec\tau\cdot\hat r
(\textstyle\frac12 {\vec\tau})^q \vec\tau\cdot\hat r
 \bar{{\cal K}}^{(j)}_{k'_\ell k'_3 s_Q} \} =  \displaystyle
\frac{(k_\ell k_3 1 q|k_\ell^\prime k_3^\prime)}{\sqrt{2 k'_\ell + 1}}
{ {^{}_{j}(k_\ell^\prime \| \vec\Theta \| k_\ell)_{i}^{}}},
\label{WE}
\end{equation}
with the ``reduced matrix elements":
\begin{equation}
\renewcommand{\arraystretch}{1.65} \begin{array}{l}
{ {^{}_{1}(k_\ell \| \vec\Theta  \| k_\ell)_{1}^{}}}
= \displaystyle\frac12\sqrt{\frac{k_\ell(2k_\ell+1)}{k_\ell+1}}, \\
{ {^{}_{2}(k_\ell \| \vec\Theta  \| k_\ell)_{2}^{}}}
= -\displaystyle\frac12\sqrt{\frac{(k_\ell+1)(2k_\ell+1)}{k_\ell}}, \\
{ {^{}_{3}(k_\ell \| \vec\Theta  \| k_\ell)_{3}^{}}}
= -\displaystyle\frac12\frac{(2k_\ell+3)
\sqrt{k_\ell}}{(k_\ell+1)(2k_\ell+1)}, \\
{ {^{}_{3}(k_\ell \| \vec\Theta  \| k_\ell)_{4}^{}}}
= -\displaystyle\frac{1}{\sqrt{2k_\ell+1}}, \\
{ {^{}_{4}(k_\ell \| \vec\Theta  \| k_\ell)_{4}^{}}}
= \displaystyle\frac12\frac{(2k_\ell-1)
\sqrt{k_\ell+1}}{\sqrt{k_\ell(2k_\ell+1)}}, \\
{ {^{}_{1}(k_\ell-1 \| \vec\Theta  \| k_\ell)_{3}^{}}}
= -\displaystyle\sqrt{\frac{(k_\ell+1)(2k_\ell-1)}{2k_\ell+1}}, \\
{ {^{}_{1}(k_\ell-1 \| \vec\Theta  \| k_\ell)_{4}^{}}}
= - \displaystyle\frac12\sqrt{\frac{(2k_\ell-1)}{k_\ell(2k_\ell+1)}},
\\
{ {^{}_{2}(k_\ell+1 \| \vec\Theta  \| k_\ell)_{3}^{}}}
= \displaystyle\frac12\sqrt{\frac{(2k_\ell+3)}{(k_\ell+1)(2k_\ell+1)}},
\\
{ {^{}_{2}(k_\ell+1 \| \vec\Theta  \| k_\ell)_{4}^{}}}
= -\displaystyle\sqrt{\frac{k_\ell(2k_\ell+3)}{2k_\ell+1}},
\end{array}
\end{equation}
and others are zero.
As far as the single-particle Fock states of the {\em same} $k_\ell$
are concerned, we can rewrite Eq.(\ref{WE}) in a more convenient form as
\begin{equation}
{^{}_a\langle k_\ell,m'| {\vec\Theta}(\infty) |k_\ell,m\rangle_b^{}}
= -c_{ab}^{} (k_\ell,m'| {{\vec K}_\ell} |k_\ell,m),
\label{WE1}
\end{equation}
where $(k_\ell,m'| {{\vec K}_\ell} |k_\ell,m)$ denotes the expectation
value of the operator ${{\vec K}_\ell}$ with respect to its eigenstates.
The multiplication coefficients analogous to the Lande's $g$-factor
can be read off from Eq.(\ref{WE}) as
$$
\renewcommand{\arraystretch}{1.5}
\begin{array}{c}
\displaystyle c_{33}^{} = +\frac{2k_\ell+3}{2(k_\ell+1)(2k_\ell+1)},
\hskip 5mm
\displaystyle c_{44}^{} = -\frac{2k_\ell-1}{2k_\ell(2k_\ell+1)}, \\
\displaystyle c_{34}^{} =
          +\frac{1}{\sqrt{k_\ell(k_\ell+1)}}\frac{1}{2k_\ell+1}, \\
\displaystyle c_{11}^{} = -\frac{1}{2(k_\ell+1)}, \hskip 5mm
\displaystyle c_{22}^{} = +\frac{1}{2k_\ell},
\end{array}\eqno(\mbox{\ref{WE1}a})
$$
and others zero. With respect to the states $|k_\ell\rangle_\pm$,
we have
$$
\renewcommand{\arraystretch}{1.5} \begin{array}{c}
\displaystyle
c_{--}^{} = 0, \hskip 5mm
c_{++}^{} = +\frac{1}{2k_\ell(k_\ell+1)}, \\
\displaystyle
c_{-+} = \frac{1}{2\sqrt{k_\ell(k_\ell+1)}}.
\end{array}\eqno(\mbox{\ref{WE1}b})
$$


\newpage

\setlength{\unitlength}{1mm}
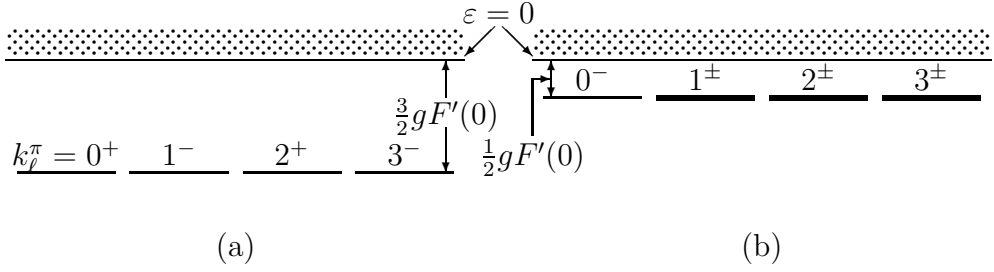
\begin{figure}
\begin{center}
\caption{Bound states of (a) heavy mesons and (b) antiflavored
heavy mesons.}
\vskip 2mm
\begin{picture}(150,42)
\thinlines
\multiput(9.5,30)(70,0){2}{\line(1,0){61}}
\put(74.5,34.5){\vector(-1,-1){4}}
\put(75.5,34.5){\vector(1,-1){4}}
\put(75,35){\makebox(0,0)[b]{$\varepsilon=0$}}
\multiput(10.75,30.75)(1.5,0){40}{\makebox(0,0){$\cdot$}}
\multiput(10,31.5)(1.5,0){41}{\makebox(0,0){$\cdot$}}
\multiput(10.75,32.25)(1.5,0){40}{\makebox(0,0){$\cdot$}}
\multiput(10,33)(1.5,0){41}{\makebox(0,0){$\cdot$}}
\multiput(10.75,33.75)(1.5,0){40}{\makebox(0,0){$\cdot$}}
\multiput(80.75,30.75)(1.5,0){40}{\makebox(0,0){$\cdot$}}
\multiput(80,31.5)(1.5,0){41}{\makebox(0,0){$\cdot$}}
\multiput(80.75,32.25)(1.5,0){40}{\makebox(0,0){$\cdot$}}
\multiput(80,33)(1.5,0){41}{\makebox(0,0){$\cdot$}}
\multiput(80.75,33.75)(1.5,0){40}{\makebox(0,0){$\cdot$}}
\put(17.5,16){\makebox(0,0)[b]{$k_\ell^\pi=0^+$}}
\put(32.5,16){\makebox(0,0)[b]{$1^-$}}
\put(47.5,16){\makebox(0,0)[b]{$2^+$}}
\put(62.5,16){\makebox(0,0)[b]{$3^-$}}
\put(68,25){\vector(0,1){5}} \put(68,21){\vector(0,-1){6}}
\put(68,22.5){\makebox(0,0){$\frac32gF'(0)$}}
\put(87.5,26){\makebox(0,0)[b]{$0^-$}}
\put(102.5,26){\makebox(0,0)[b]{$1^\pm$}}
\put(117.5,26){\makebox(0,0)[b]{$2^\pm$}}
\put(132.5,26){\makebox(0,0)[b]{$3^\pm$}}
\put(82,25){\vector(0,1){5}} \put(82,30){\vector(0,-1){5}}
\put(79.5,20){\line(0,1){7.5}} \put(79.5,27.5){\vector(1,0){2.5}}
\put(79.5,19.5){\makebox(0,0)[t]{$\frac12gF'(0)$}}
\put(40,5){\makebox(0,0){(a)}} \put(110,5){\makebox(0,0){(b)}}
\thicklines
\multiput(11,15)(15,0){4}{\line(1,0){13}}
\put(81,25){\line(1,0){13}}
\multiput(96,25.2)(15,0){3}{\line(1,0){13}}
\multiput(96,25.0)(15,0){3}{\line(1,0){13}}
\multiput(96,24.8)(15,0){3}{\line(1,0){13}}
\end{picture}
\end{center}
\end{figure}

\setlength{\unitlength}{1mm}
\begin{figure}
\begin{center}
\caption{Heavy baryon spectrum.}
\vskip 2mm
\begin{picture}(150,38)
\thinlines
\put(35,30){\line(1,0){80}}
\put(75,32){\makebox(0,0)[b]{$m_N + m_\Phi$}}
\put(40.5,9){\makebox(0,0)[r]{$i$=0,$j$=1/2}}
\put(41,9){\line(1,0){4}}
\put(40.5,16.5){\makebox(0,0)[r]{$i$=1,$j$=1/2}}
\put(41,16.5){\line(2,1){4}}
\put(40.5,20.5){\makebox(0,0)[r]{$i$=1,$j$=3/2}}
\put(41,20.5){\line(2,-1){4}}
\put(109.5,7){\makebox(0,0)[l]{$i$=0,$j$=1/2}}
\put(109,7){\line(-2,1){4}}
\put(109.5,11){\makebox(0,0)[l]{$i$=0,$j$=3/2}}
\put(109,11){\line(-2,-1){4}}
\put(109.5,16.5){\makebox(0,0)[l]{$i$=1,$j$=1/2}}
\put(109,16.5){\line(-2,1){4}}
\put(109.5,20.5){\makebox(0,0)[l]{$i$=1,$j$=3/2}}
\put(109,20.5){\line(-2,-1){4}}
\put(55,10){\makebox(0,0)[b]{$\Lambda_Q$}}
\put(55,20.5){\makebox(0,0)[b]{$\Sigma_Q,\Sigma^*_Q$}}
\put(55,0){\makebox(0,0){$\pi=+$}}
\put(95,0){\makebox(0,0){$\pi=-$}}
\put(71,9){\line(1,0){3}}
\put(72.5,21.5){\vector(0,1){8.5}} \put(72.5,17.5){\vector(0,-1){8.5}}
\put(72.5,19.5){\makebox(0,0){$\frac32gF'(0)$}}
\multiput(81,9)(0,9.5){2}{\line(1,0){3}}
\put(82.5,15.5){\vector(0,1){3}} \put(82.5,12){\vector(0,-1){3}}
\put(82.5,13.5){\makebox(0,0){$1/{\cal I}$}}
\multiput(55,26)(40,0){2}{\makebox(0,0){$\vdots$}}
\thicklines
\multiput(45,9)(40,0){2}{\line(1,0){20}}
\multiput(45,18.5)(40,0){2}{\line(1,0){20}}
\end{picture}
\end{center}
\end{figure}
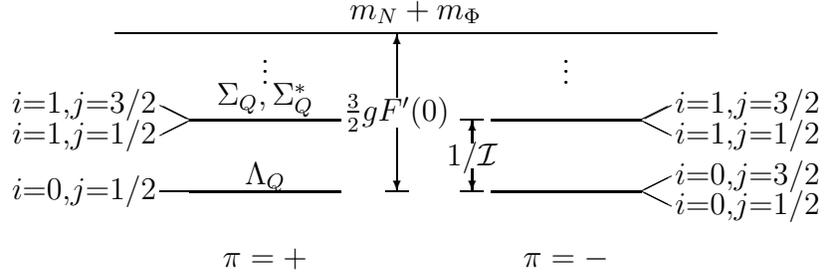

\setlength{\unitlength}{1mm}
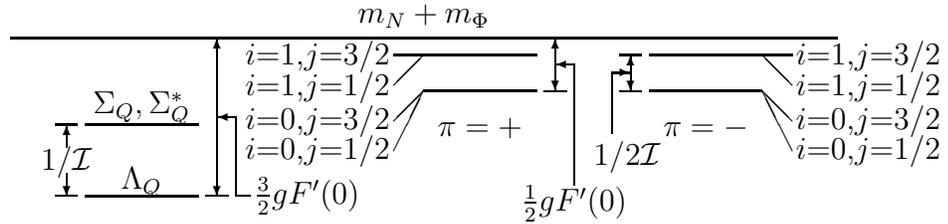
\begin{figure}
\begin{center}
\caption{P-baryon spectrum.}
\vskip 2mm
\begin{picture}(150,38)
\thinlines
\put(20,30){\line(1,0){110}}
\put(75,32){\makebox(0,0)[b]{$m_N + m_\Phi$}}
\put(37.5,10){\makebox(0,0)[b]{$\Lambda_Q$}}
\put(37.5,20){\makebox(0,0)[b]{$\Sigma_Q,\Sigma^*_Q$}}
\put(46,9){\line(1,0){3}}
\put(47.5,19.5){\vector(0,1){10.5}}
\put(47.5,19.5){\vector(0,-1){10.5}}
\put(50,19.5){\vector(-1,0){2.5}}
\put(50,19.5){\line(0,-1){10.5}}
\put(50,9){\line(1,0){2}}
\put(52.5,9){\makebox(0,0)[l]{$\frac32gF'(0)$}}
\multiput(26,9)(0,9.5){2}{\line(1,0){3}}
\put(27.5,15.5){\vector(0,1){3}}
\put(27.5,12){\vector(0,-1){3}}
\put(27.5,13.5){\makebox(0,0){$1/{\cal I}$}}
\put(70.5,23.5){\makebox(0,0)[r]{$i$=1,$j$=1/2}}
\put(71,23.75){\line(1,1){4}}
\put(70.5,27.5){\makebox(0,0)[r]{$i$=1,$j$=3/2}}
\put(71,27.75){\line(1,0){4}}
\put(70.5,19){\makebox(0,0)[r]{$i$=0,$j$=3/2}}
\put(71,19){\line(1,1){4}}
\put(70.5,15){\makebox(0,0)[r]{$i$=0,$j$=1/2}}
\put(71,15){\line(1,2){4}}
\put(124.5,23.5){\makebox(0,0)[l]{$i$=1,$j$=1/2}}
\put(124,23.75){\line(-1,1){4}}
\put(124.5,27.5){\makebox(0,0)[l]{$i$=1,$j$=3/2}}
\put(124,27.75){\line(-1,0){4}}
\put(124.5,19){\makebox(0,0)[l]{$i$=0,$j$=3/2}}
\put(124,19){\line(-1,1){4}}
\put(124.5,15){\makebox(0,0)[l]{$i$=0,$j$=1/2}}
\put(124,15){\line(-1,2){4}}
\multiput(101,23)(0,4.75){2}{\line(1,0){3}}
\put(102.5,23){\vector(0,1){4.75}}
\put(102.5,27.75){\vector(0,-1){4.75}}
\put(100,25.375){\vector(1,0){2.5}}
\put(100,25.375){\line(0,-1){8}}
\put(102,16.375){\makebox(0,0)[t]{$1/2{\cal I}$}}
\put(91,23){\line(1,0){3}}
\put(92.5,26.5){\vector(0,1){3.5}}
\put(92.5,26.5){\vector(0,-1){3.5}}
\put(95,26.5){\vector(-1,0){2.5}}
\put(95,26.5){\line(0,-1){16}}
\put(95,10.5){\makebox(0,0)[t]{$\frac12gF'(0)$}}
\put(82.5,18){\makebox(0,0){$\pi=+$}}
\put(112.5,18){\makebox(0,0){$\pi=-$}}
\thicklines
\put(30,9){\line(1,0){15}}
\put(30,18.5){\line(1,0){15}}
\multiput(75,23)(30,0){2}{\line(1,0){15}}
\multiput(75,27.75)(30,0){2}{\line(1,0){15}}
\end{picture}
\end{center}
\end{figure}

\begin{table}
\begin{center}
\caption{Heavy-meson eigenstates  with $\ell=0$ and 1.}
\vskip 2mm
\begin{tabular}{cccccl}
$\{ k_\ell, k_3,\pi,s^{}_Q\}$  & $\varepsilon$ & eigenfunct. &
$\ell_{\mbox{\scriptsize eff}}$  & $k$ & $|n\rangle$ \\
\hline
$\{0,0,-,s^{}_Q\}$ & $+\frac12gF^\prime(0)$ &
$f(r){\cal K}^{(1)}_{00s^{}_Q}$ & 1 & $\frac12$ & $|0\rangle_1$ \\
$\{1,m,-,s^{}_Q\}$ & $-\frac32gF^\prime(0)$ &
$f(r){\cal K}^{(-)}_{1ms^{}_Q}$ & 1 &
$\frac12,\frac32$ & $|1\rangle_{-}$ \\
$\{1,m,-,s^{}_Q\}$ & $+\frac12gF^\prime(0)$ &
$f(r){\cal K}^{(+)}_{1ms^{}_Q}$ & 1 &
$\frac12,\frac32$ & $|1\rangle_{+}$ \\
\hline
$\{0,0,+,s^{}_Q\}$ & $-\frac32gF^\prime(0)$ &
$f(r){\cal K}^{(3)}_{00s^{}_Q}$ & 0 &
$\textstyle\frac12 $ & $|0\rangle_3$ \\
$\{1,m,+,s^{}_Q\}$ & $+\frac12gF^\prime(0)$ &
 $f(r){\cal K}^{(1)}_{1ms^{}_Q}$ & 2 &
$\textstyle\frac12 ,\frac32$ & $|1\rangle_1$ \\
$\{1,m,+,s^{}_Q\}$ & $+\frac12gF^\prime(0)$ &
 $f(r){\cal K}^{(2)}_{1ms^{}_Q}$ & 0 &
$\textstyle\frac12 ,\frac32$ & $|1\rangle_2$ \\
$\{2,m,+,s^{}_Q\}$ & $-\frac32gF^\prime(0)$ &
 $f(r){\cal K}^{(-)}_{2ms^{}_Q}$ & 2 &
$\frac32,\frac52$ & $|2\rangle_{-}$ \\
$\{2,m,+,s^{}_Q\}$ & $+\frac12gF^\prime(0)$ &
 $f(r){\cal K}^{(+)}_{2ms^{}_Q}$ & 2 &
$\frac32,\frac52$ & $|2\rangle_{+}$ \\
\end{tabular}
\end{center}
\end{table} 

\begin{table}
\begin{center}\renewcommand{\arraystretch}{1.5}
\caption{$|i,j_\ell^\pi\rangle\!\rangle$ states for heavy baryons.}
\vskip 3mm
\begin{tabular}{ccllccc}
 $i$ & $j_\ell^\pi$ &
 $\displaystyle\renewcommand{\arraystretch}{1} \begin{array}{l}
 |n\rangle \end{array}$ &
$\displaystyle\renewcommand{\arraystretch}{1} \begin{array}{l}
 |i,j^\pi_\ell\rangle\!\rangle_a^{} \end{array}$ &
$\displaystyle\renewcommand{\arraystretch}{1} \begin{array}{c}
 \varepsilon \end{array}$ & $j$ & \\ \hline
\hline 
0 & $0^+$ &
$\displaystyle\renewcommand{\arraystretch}{1} \begin{array}{l}
 |0\rangle \end{array}$ &
 $\displaystyle\renewcommand{\arraystretch}{1} \begin{array}{l}
 |0,0^+\rangle\!\rangle \end{array}$ &
$\displaystyle\renewcommand{\arraystretch}{1} \begin{array}{c}
 -\frac32gF^\prime(0) \end{array}$ & $\frac12$ & $\Lambda_Q$ \\
\hline 
1 & $1^+$ &
$\displaystyle\renewcommand{\arraystretch}{1} \begin{array}{l}
 |0\rangle_3  \\  |2\rangle_-  \end{array}$ &
$\displaystyle\renewcommand{\arraystretch}{1} \begin{array}{l}
 |1,1^+\rangle\!\rangle_1 \\ |1,1^+\rangle\!\rangle_2 \end{array}$ &
$\displaystyle\renewcommand{\arraystretch}{1} \begin{array}{c}
 -\frac32gF^\prime(0) \\ -\frac32gF^\prime(0) \end{array}$ &
$\textstyle\frac12 ,\frac32$ & $\Sigma_Q,\,\Sigma^*_Q$ \\ \hline
\hline 
0 & $1^-$ &
$\displaystyle\renewcommand{\arraystretch}{1} \begin{array}{l}
 |1\rangle_-  \end{array}$ &
$\displaystyle\renewcommand{\arraystretch}{1} \begin{array}{l}
 |0,1^-\rangle\!\rangle  \end{array}$ &
$\displaystyle\renewcommand{\arraystretch}{1} \begin{array}{c}
 -\frac32gF^\prime(0)  \end{array}$ &
$\frac12,\frac32$ &  \\
\hline 
1 & $0^-$ &
$\displaystyle\renewcommand{\arraystretch}{1} \begin{array}{l}
 |1\rangle_- \end{array}$ &
$\displaystyle\renewcommand{\arraystretch}{1} \begin{array}{l}
 |1,0^-\rangle\!\rangle \end{array}$ &
$\displaystyle\renewcommand{\arraystretch}{1} \begin{array}{c}
 -\frac32gF^\prime(0) \end{array}$ &
$\frac12$ &  \\
\hline 
1 & $1^-$ &
$\displaystyle\renewcommand{\arraystretch}{1} \begin{array}{l}
 |1\rangle_- \end{array}$ &
$\displaystyle\renewcommand{\arraystretch}{1} \begin{array}{l}
 |1,1^-\rangle\!\rangle \end{array}$ &
$\displaystyle\renewcommand{\arraystretch}{1} \begin{array}{c}
 -\frac32gF^\prime(0) \end{array}$ &
$\frac12,\frac32$ & \\
\end{tabular}
\end{center}
\end{table}

\begin{table}
\begin{center}\renewcommand{\arraystretch}{1.5}
\caption{$|i,j^\pi_\ell\rangle\!\rangle$ states for the $P$-baryons.}
\vskip 3mm
\begin{tabular}{ccllcc}
 $i$ & $j_\ell^\pi$ & $\displaystyle\renewcommand{\arraystretch}{1}
\begin{array}{l} |n\rangle \end{array}$ &
$\displaystyle\renewcommand{\arraystretch}{1} \begin{array}{l}
 |i,j^\pi_\ell\rangle\!\rangle_i \end{array}$ & $\varepsilon$
& $j$ \\ \hline
\hline 
 0 & $0^-$ & $\displaystyle\renewcommand{\arraystretch}{1}
 \begin{array}{l} |0\rangle_1 \end{array}$ &
$\displaystyle\renewcommand{\arraystretch}{1} \begin{array}{l}
 |0,0^-\rangle\!\rangle \end{array}$ & $-\frac12gF^\prime(0)$ &
 $\textstyle\frac12 $   \\
\hline 
 0 & $1^-$ & $\displaystyle\renewcommand{\arraystretch}{1}
\begin{array}{l} |1\rangle_+ \end{array}$ &
$\displaystyle\renewcommand{\arraystretch}{1} \begin{array}{l}
 |0,1^-\rangle\!\rangle \end{array}$ & $-\frac12gF^\prime(0)$ &
 $\textstyle\frac12 ,\frac32$  \\
\hline 
 1 & $0^-$ & $\displaystyle\renewcommand{\arraystretch}{1}
 \begin{array}{l} |1\rangle_+ \end{array}$ &
$\displaystyle\renewcommand{\arraystretch}{1} \begin{array}{l}
 |0,1^-\rangle\!\rangle \end{array}$ & $-\frac12gF^\prime(0)$ &
 $\textstyle\frac12 $  \\
\hline 
 1 & $1^-$ & $\displaystyle\renewcommand{\arraystretch}{1}
 \begin{array}{l} |0\rangle_1 \\ |1\rangle_+ \\
 |2\rangle_1 \\ |2\rangle_2 \end{array}$ &
$\displaystyle\renewcommand{\arraystretch}{1} \begin{array}{l}
 |1,1^-\rangle\!\rangle_1 \\ |1,1^-\rangle\!\rangle_2 \\
|1,1^-\rangle\!\rangle_3 \\ |1,1^-\rangle\!\rangle_4 \end{array}$
 & $-\frac12gF^\prime(0)$ &
$\textstyle\frac12 ,\frac32$  \\ \hline
\hline 
 0 & $1^+$ &
$\displaystyle\renewcommand{\arraystretch}{1} \begin{array}{l}
 |1\rangle_1 \\ |1\rangle_2 \end{array}$ &
$\displaystyle\renewcommand{\arraystretch}{1} \begin{array}{l}
 |0,1^+\rangle\!\rangle_1 \\ |0,1^+\rangle\!\rangle_2 \end{array}$
& $-\frac12gF^\prime(0)$ & $\textstyle\frac12 ,\frac32$  \\
\hline 
 1 & $0^+$ &
$\displaystyle\renewcommand{\arraystretch}{1} \begin{array}{l}
 |1\rangle_1 \\ |1\rangle_2 \end{array}$ &
$\displaystyle\renewcommand{\arraystretch}{1} \begin{array}{l}
 |1,0^+\rangle\!\rangle_1 \\ |1,0^+\rangle\!\rangle_2 \end{array}$
& $-\frac12gF^\prime(0)$ & $\textstyle\frac12 $  \\
\hline 
 1 & $1^+$ &
$\displaystyle\renewcommand{\arraystretch}{1} \begin{array}{l}
 |1\rangle_1 \\ |1\rangle_2 \\ |2\rangle_+ \end{array}$ &
$\displaystyle\renewcommand{\arraystretch}{1} \begin{array}{l}
 |1,1^+\rangle\!\rangle_1 \\ |1,1^+\rangle\!\rangle_2 \\
 |1,1^+\rangle\!\rangle_3 \end{array}$
& $-\frac12gF^\prime(0)$ & $\textstyle\frac12 ,\,\frac32$  \\
\end{tabular}
\end{center}
\end{table}

\begin{table} 
\begin{center}\renewcommand{\arraystretch}{1.5}
\caption{Positive and Negative Parity $P$-baryon (P) masses (in MeV).}
\vskip 3mm
\begin{tabular}{ccccccr}
 $i$ & $j_\ell^\pi$ & $j^\pi$ & Mass Formula &
 $m^{}_{P_{\bar{c}}^{}}$ & $m^{}_{P_{\bar{b}}^{}}$
& \multicolumn{1}{c}{b.e.$^*$} \\
\hline
 0 & $0^-$ & ${\frac12}^-$ &
 $M_{sol} + \omega'_B + 3/8{\cal I}$ & 2704 & 6042 & 210 \\
 0 & $1^-$ & ${\frac12}^-,{\frac32}^-$ &
 $M_{sol} + \omega'_B + 3/8{\cal I}$ & 2704 & 6042 & 210 \\
 1 & $1^-$ & ${\frac12}^-,{\frac32}^-$ &
 $M_{sol} + \omega'_B + 7/8{\cal I}$ & 2802 & 6140 & 112 \\
\hline
 0 & $1^+$ & ${\frac12}^+,{\frac32}^+$ &
 $M_{sol} + \omega'_B + 3/8{\cal I}$ &  2704 & 6042 & 210 \\
 1 & $0^+$ & ${\frac12}^+$ &
 $M_{sol} + \omega'_B + 7/8{\cal I}$ &  2802 & 6140 & 112 \\
 1 & $1^+$ & ${\frac12}^+,{\frac32}^+$ &
 $M_{sol} + \omega'_B + 7/8{\cal I}$ &  2802 & 6140 & 112 \\
\end{tabular} \end{center}
{$^*$ Binding energy below $M_{th}$.}
\end{table} 


\begin{references}
\bibitem{fn1}
Actually, in case of $k_\ell=0$, we have zero expectation value
of the ${\vec\Theta}(\infty)$ and $c$ cannot be a well-defined
quantity. For simplicity, we take its value as zero. The final
results do not depend on it, anyway. As for the antiflavored
heavy meson bound states, since the operator ${\vec\Theta}$ is
defined with a multiplication factor $v_0$, $c$-values should
change their sign.
\end{references}
\end{document}